\documentclass[twocolumn]{aastex631}
\usepackage{hhline}
\usepackage{xspace}
\usepackage{natbib}
\usepackage{bigstrut}
\usepackage{lipsum} 
\usepackage[T1]{fontenc}

\usepackage{graphicx}
\usepackage{amsmath}
\usepackage{xcolor}
\usepackage{bm}

\newcommand{\RXTE}{\textit{RXTE}\xspace}

\shorttitle{Variability and state transitions in GX 339-4}
\shortauthors{Lucchini et al.}
\graphicspath{{./}{figures/}}

\begin{document}

\title{Variability as a predictor for the hard-to-soft state transition in GX 339-4}

\author[0000-0002-2235-3347]{Matteo Lucchini}
\affiliation{MIT Kavli Institute for Astrophysics and Space Research, MIT, 70 Vassar Street, Cambridge, MA 02139, USA}

\author{Marina Ten Have}
\affiliation{MIT Kavli Institute for Astrophysics and Space Research, MIT, 70 Vassar Street, Cambridge, MA 02139, USA}

\author{Jingyi Wang}
\affiliation{MIT Kavli Institute for Astrophysics and Space Research, MIT, 70 Vassar Street, Cambridge, MA 02139, USA}

\author{Jeroen Homan}
\affiliation{Eureka Scientiﬁc, Inc., 2452 Delmer Street, Oakland, CA 94602, USA}
\affiliation{SRON, Netherlands Institute for Space Research, Sorbonnelaan 2, 3584 CA Utrecht, The Netherlands}

\author{Erin Kara}
\affiliation{MIT Kavli Institute for Astrophysics and Space Research, MIT, 70 Vassar Street, Cambridge, MA 02139, USA}

\author{Oluwashina Adegoke}
\affiliation{Cahill Center for Astronomy and Astrophysics, California Institute of Technology, Pasadena, CA 91125, USA}

\author{Riley Connors}
\affiliation{Villanova University, Villanova, PA 19085, USA}

\author{Thomas Dauser}
\affiliation{Dr. Karl Remeis-Observatory and Erlangen Centre for Astroparticle Physics, Friedrich-Alexander-Universit\"at Erlangen-N\"urnberg, Sternwartstr.~7, 96049 Bamberg, Germany}

\author{Javier Garcia}
\affiliation{Cahill Center for Astronomy and Astrophysics, California Institute of Technology, Pasadena, CA 91125, USA}

\author{Guglielmo Mastroserio}
\affiliation{INAF-Osservatorio Astronomico di Cagliari, via della Scienza 5, I-09047 Selargius (CA), Italy}

\author{Adam Ingram}
\affiliation{Department of Physics, Astrophysics, University of Oxford, Denys Wilkinson Building, Keble Road, Oxford OX1 3RH, UK}
\affiliation{School of Mathematics, Statistics and Physics, Newcastle University, Herschel Building, Newcastle upon Tyne, NE1 7RU, UK}

\author{Michiel van der Klis}
\affiliation{Anton Pannekoek Institute for Astronomy, University of Amsterdam, Science Park 904, NL-1098 XH Amsterdam, Netherlands}

\author{Ole K\"{o}nig}
\affiliation{Dr. Karl Remeis-Observatory and Erlangen Centre for Astroparticle Physics, Friedrich-Alexander-Universit\"at Erlangen-N\"urnberg, Sternwartstr.~7, 96049 Bamberg, Germany}

\author{Collin Lewin}
\affiliation{MIT Kavli Institute for Astrophysics and Space Research, MIT, 70 Vassar Street, Cambridge, MA 02139, USA}

\author{Labani Mallick}
\affiliation{Cahill Center for Astronomy and Astrophysics, California Institute of Technology, Pasadena, CA 91125, USA}
\affiliation{University of Manitoba, Department of Physics \& Astronomy, Winnipeg, Manitoba R3T 2N2, Canada}
\affiliation{Canadian Institute for Theoretical Astrophysics, University of Toronto, 60 St George Street, Toronto, Ontario M5S 3H8, Canada}

\author{Edward Nathan}
\affiliation{Department of Physics, Astrophysics, University of Oxford, Denys Wilkinson Building, Keble Road, Oxford OX1 3RH, UK}
\affiliation{Cahill Center for Astronomy and Astrophysics, California Institute of Technology, Pasadena, CA 91125, USA}

\author{Patrick O'Neill}
\affiliation{School of Mathematics, Statistics and Physics, Newcastle University, Herschel Building, Newcastle upon Tyne, NE1 7RU, UK}

\author{Christos Panagiotou}
\affiliation{MIT Kavli Institute for Astrophysics and Space Research, MIT, 70 Vassar Street, Cambridge, MA 02139, USA}

\author{Joanna Piotrowska}
\affiliation{Cahill Center for Astronomy and Astrophysics, California Institute of Technology, Pasadena, CA 91125, USA}

\author{Phil Uttley}
\affiliation{Anton Pannekoek Institute for Astronomy, University of Amsterdam, Science Park 904, NL-1098 XH Amsterdam, Netherlands}

\begin{abstract}

During the outbursts of black hole X-ray binaries (BHXRBs), their accretion flows transition through several states. The source luminosity rises in the hard state, dominated by non-thermal emission, before transitioning to the blackbody-dominated soft state. As the luminosity decreases, the source transitions back into the hard state and fades to quiescence. This picture does not always hold, as $\approx40\%$ of the outbursts never leave the hard state. Identifying the physics that govern state transitions remains one of the outstanding open questions in black hole astrophysics.  In this paper we present an analysis of archival RXTE data of multiple outbursts of GX 339-4. We compare the properties of the X-ray variability and time-averaged energy spectrum and demonstrate that the variability (quantified by the power spectral hue) systematically evolves $\approx10-40$ days ahead of the canonical state transition (quantified by a change in spectral hardness); no such evolution is found in hard state only outbursts. This indicates that the X-ray variability can be used to predict if and when the hard-to-soft state transition will occur. Finally, we find a similar behavior in ten outbursts of four additional BHXRBs with more sparse observational coverage. Based on these findings, we suggest that state transitions in BHXRBs might be driven by a change in the turbulence in the outer regions of the disk, leading to a dramatic change in variability. This change is only seen in the spectrum days to weeks later, as the fluctuations propagate inwards towards the corona. 

\end{abstract}

\section{Introduction} \label{sec:intro}

X-ray binaries (XRBs) are systems in which a stellar mass compact object accretes from a companion star. XRBs are further classified as low or high mass systems depending on the mass $M_{*}$ of the companion ($M_{*}\approx\,1$ and $10\,M_{\odot}$, respectively, e.g. \citealt{Tauris06}), and as neutron star or black hole systems depending on the nature of the compact object. Most low mass black hole XRBs (BHXRBs) are transient sources: they undergo bright outbursts on timescales of $\approx\,$months, in between quiescent periods lasting $\approx\,$years or more \citep[e.g.][]{Tetarenko16}.

During each outburst, the X-ray spectrum of a BHXRB is dominated by two components: a thermal component, peaking in the soft X-rays around $1\,\rm{keV}$, and a non-thermal power-law component extending up to tens or hundreds of $\rm{keV}$. The former originates from a geometrically thin, optically thick accretion disk \citep[e.g.][]{Shakura73}, while the latter is powered by an optically thin ($\tau\approx1$), hot ($kT_{\rm e}\approx100\,\rm{keV}$) plasma called the corona, which inverse Compton scatters disk photons up to hard X-ray energies \citep[e.g.][]{Shapiro76}. 

The majority of BHXRB outbursts display a consistent behavior \citep[e.g.][]{Homan05,Remillard06,Belloni10}, which is typically quantified by plotting spectral hardness (defined here as the ratio of count rates in a hard and soft X-ray band) against intensity (or luminosity). At the start of an outburst, the corona dominates the X-ray spectrum, and the X-ray emission is highly variable, as the source rises in luminosity through the hard state (HS). This continues until the source reaches a significant fraction ($\approx10\%$) of the Eddington luminosity $L_{\rm Edd}$; at this point, the corona weakens and the disk begins to dominate the emission, causing the X-ray spectrum to soften rapidly. This state is referred to as the ``intermediate state'' (IMS, further divided between hard and soft intermediate state, or HIMS and SIMS), and the evolution continues until the corona disappears almost entirely, at which point the source has reached the ``soft state'' (SS). After reaching the SS, the source decreases in luminosity until $\approx0.1-1\%L_{\rm Edd}$, at which point it transitions back to the HS and then fades into quiescence. This trend forms a `q' shape in the hardness-intensity diagram (HID, e.g. \citealt{Homan01}). Different outflows are launched from the system depending on the state. During the HS and IMS, synchrotron-emitting jets are observed ubiquitously \citep[e.g.][]{Fender04a,Fender04b}, while in the SS the jet is quenched but isotropic winds are common \citep[e.g.][]{Ponti12,DiazTrigo16}. 

The properties of the X-ray variability also change as a function of state \citep[e.g.][]{Remillard06,Belloni10}. Early in the HS, the light curves display featureless flat-top noise across a broad range of Fourier frequencies, as well as large ($\geq 10\%$) rms. As the source transitions to the HIMS, type-C QPOs appear and the broadband noise becomes band-limited at $\approx$ a few Hz. Eventually, the broad band noise disappears almost completely and the type-C QPO is replaced by a type-B QPO, signalling the onset of the SIMS. The SS shows flat-top noise of similar shape to the HS, but with much lower ($\leq0.1\%$) rms. Finally, two sources (GRS 1915+105 and IGR J17091-3624) show multiple classes with unique variability signatures that do not fit this standard picture \citep[e.g.][]{Belloni00,Altamirano11,Court17}; these are sometimes referred to as the ``heartbeat'' sources.

Decades after its discovery \citep{Tananbaum72}, the exact physics that drive the state transition are poorly understood. This is in part because the state transition occurs very rapidly (e.g. \citealt{Remillard06,Tetarenko16}, this work), and therefore is often the least constrained observationally. Furthermore, roughly 40\% of outbursts never reach the SS, instead remaining in either the HS or IMS \citep{Tetarenko16}. These are typically referred to as ``failed'' outbursts, or ``hard-state'' outbursts \citep[e.g.][]{Belloni02,Capitanio09,Bassi19}; in this paper we adopt the latter name. Recently, several authors have attempted to identify a diagnostic that can be used during the rising hard state, in order to predict when, if at all, a given outburst will result in a state transition. Finding such a diagnostic would provide invaluable insight into the mechanism(s) that regulate state transitions and, more broadly, the evolution of accretion flows as a function of accretion rate. Unfortunately, results have been somewhat inconclusive. \cite{Furst15} and \cite{deHaas21} showed that GX 339$-$4 follows different tracks on the radio/X-ray plane (\citealt{Hannikainen98}, \citealt{Corbel00}) during full and hard state outbursts, indicating that the coupling between disk and jet may be different depending on the type of outburst. These authors also suggested that the radio/X-ray correlation could be used in the early phases of the outburst to predict its outcome. However, this behavior may not be universal: H1743$-$322, for instance, shows the same radio/X-ray correlation in full and hard state outbursts \citep{Williams20}. To date, the most comprehensive comparison of the properties of full and hard state outbursts was published by \cite{Alabarta21}. These authors found that X-ray properties (like HID, variability or light-curve shape) alone are not sufficient to predict whether or when the state transition will occur. Additionally, these authors found that GX 339$-$4 is brighter in the O/IR bands before a hard state outburst, compared to a full one. They suggested that the process responsible for triggering the state transition is connected to the properties of the disk in quiescence - either the size of the emitting region, or the accretion rate, should be higher before hard state outbursts, compared to full ones. Finally, \cite{Kalemci04} found that during the decay of an outburst, the X-ray rms changes ahead of the X-ray spectral properties, and state transitions are more easily defined by changes in the former, rather than the latter.

In this paper we present a re-analysis of archival \RXTE data and demonstrate that in GX 339$-$4, the X-ray variability during the rise can indeed be used to predict the outcome of an outburst. In particular, the power spectral hue begins its transitions before the ``canonical'' state transition; this evolution never occurs during hard state outbursts. We refer the reader to \cite{Heil15a,Heil15b} for more details on the power spectral hue, but briefly, it is an estimate of the shape of the X-ray power spectrum, independent of its normalization. This change in power spectral hue is mainly caused by a decrease in the low-frequency broadband noise $\approx2-5$ weeks ahead of the state transition. We find that the behavior of ten outbursts from four additional sources (GRO 1655$-$40, XTE J1550$-$564, XTE J1752$-$223 and H1743$-$322) is consistent with that of GX 339$-$4, although in these sources the observational coverage is more limited.

The paper is structured as follows. In Sec.\ref{sec:data_reduction} we present the details of the data reduction process, in Sec.\ref{sec:results} we present the main results of our analysis, and in Sec.\ref{sec:discussion} we discuss these results and draw our conclusions. 

\section{Data reduction}\label{sec:data_reduction}

We focus mainly on seven outbursts of GX 339$-$4 covered by \RXTE between 2002 and 2010. The source went into a further outburst in 1998, but unfortunately only a few \RXTE observations were taken during the rising hard state, and therefore we exclude it from this work. We also selected a sample of outbursts from the sources analyzed in \cite{Heil15a} with observations starting in the hard state and good \RXTE coverage up to the state transition (or the decay, for hard state outbursts). These are the 2005 outburst of GRO 1655$-$40, the 2009 outburst of XTE J1752$-$223, the 2008, 2010 and 2011 outbursts of H1743$-$322, and all five  outbursts of XTE J1550-564 (1999, 2000, 2001, 2002 and 2003). The properties of the outbursts analyzed in this work are summarized in the Appendix in Tab.\ref{tab:outburst_date_1} and Tab.\ref{tab:outburst_date_2}. In this paper, we concentrate on BH transients showing ``canonical'' outburst behavior \cite[e.g][]{Remillard06,Alabarta21} and do not consider the two ``heartbeat'' sources whose behavior is more complex.
\begin{figure}
    \centering
    \includegraphics[width=\columnwidth, trim={0.65cm 0.0cm 0.6cm 0.0cm},clip]{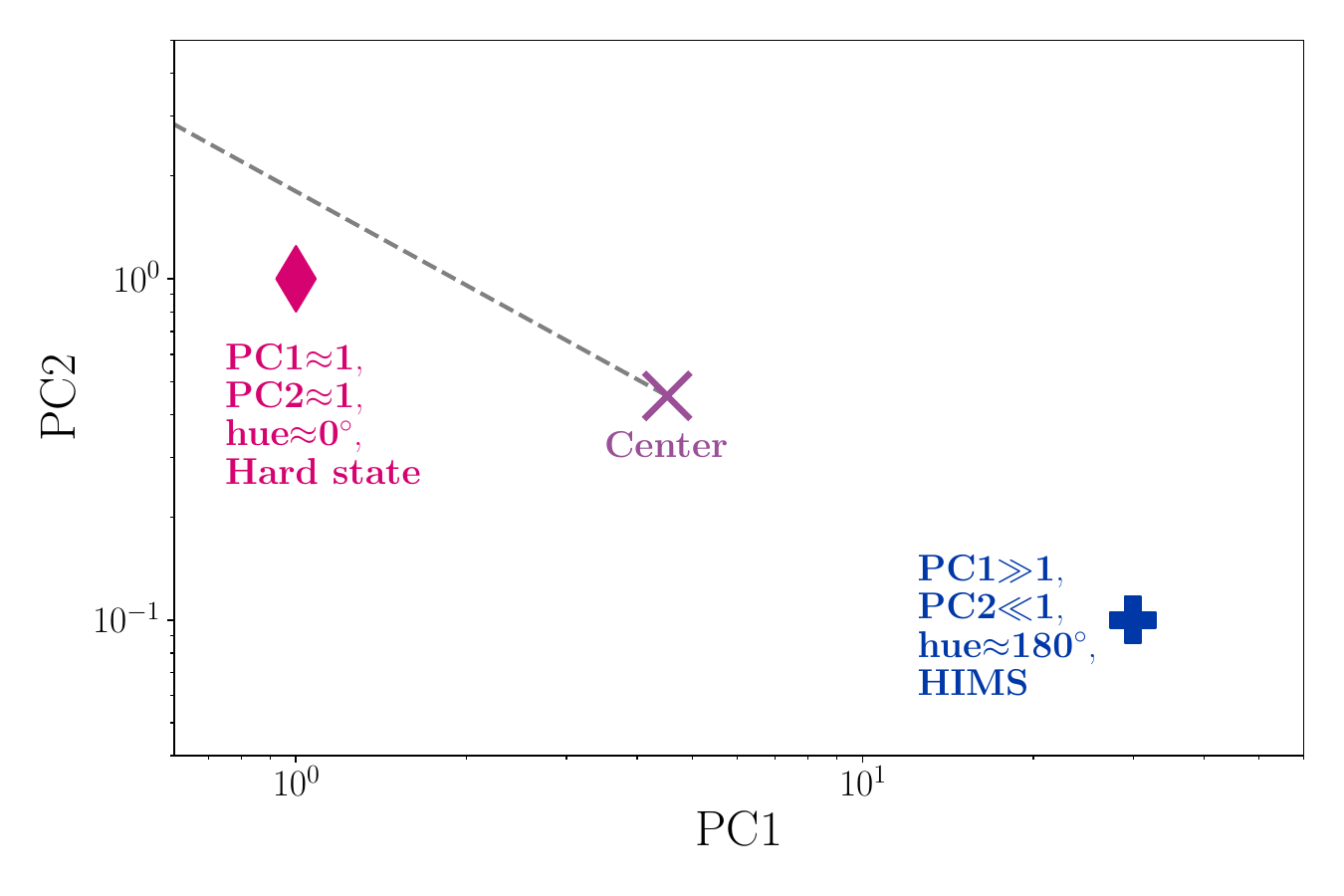}
    \caption{Example of the hue evolution. The purple cross indicates the central point at (PC1,PC2)=(4.51920, 0.453724), and the dashed gray line the line is the reference from which the hue angle is calculated. The scarlet diamond, near (PC1,PC2)=(1,1), corresponds to the low HS. The blue cross, at (PC1,PC2)=(30,0.1), corresponds to the HIMS.}
    \label{fig:hueexample}
\end{figure}
\begin{figure*}
    \centering
    \includegraphics[width=\columnwidth, trim={0.65cm 0.0cm 0.6cm 0.0cm},clip]{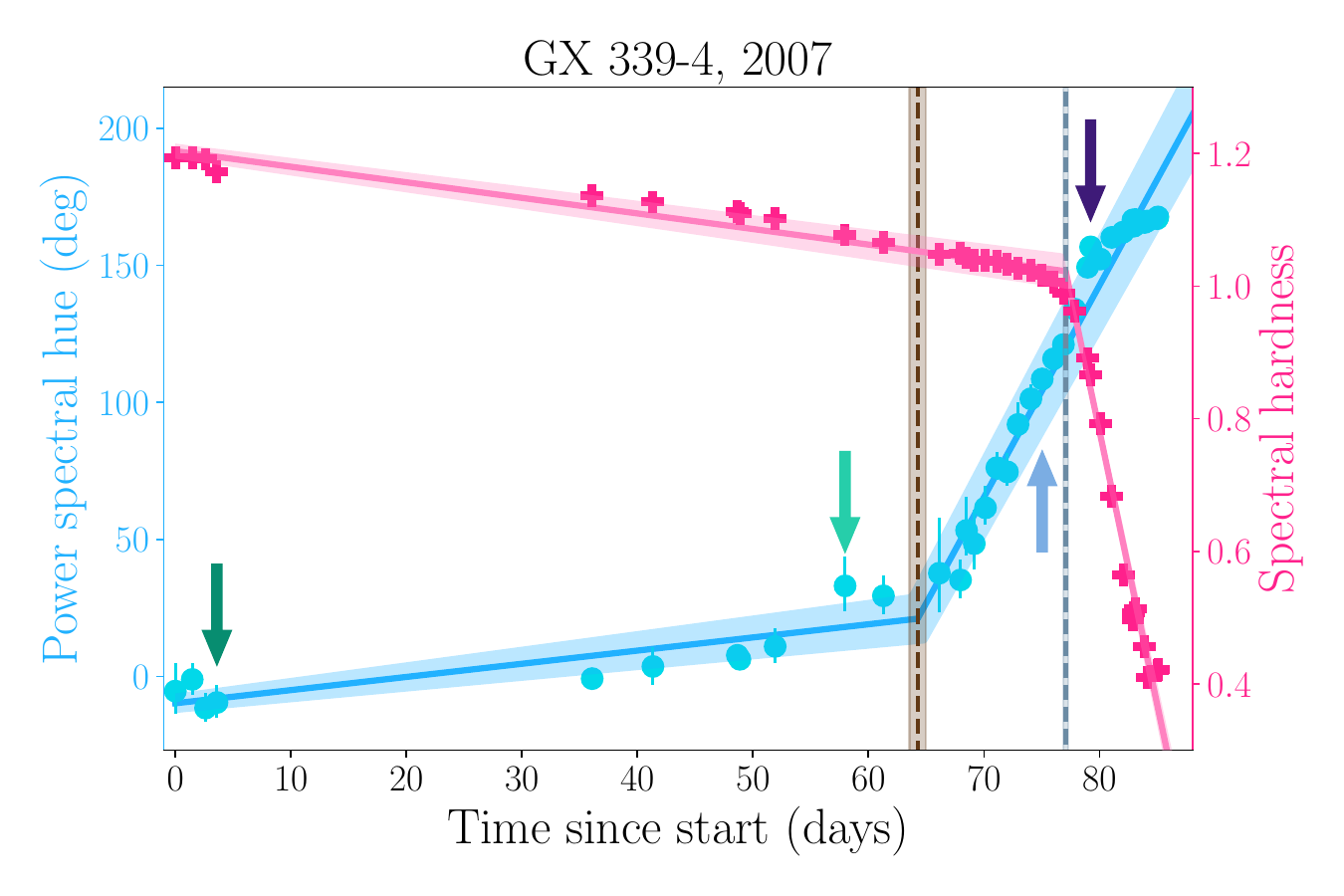}
    \includegraphics[width=\columnwidth, trim={0.65cm 0.0cm 0.6cm 0.0cm},clip]{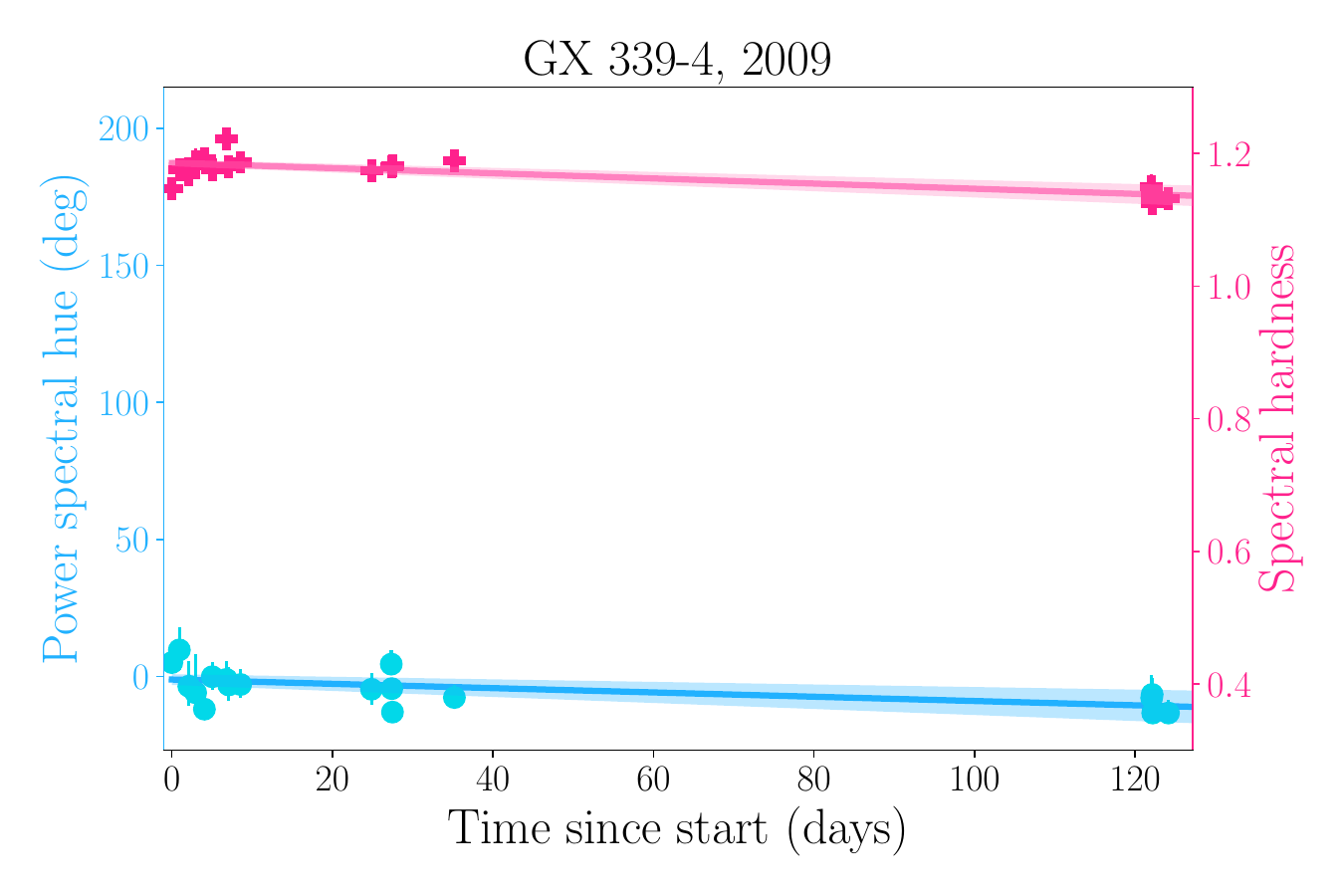}
    \caption{Left panel: evolution over time of the power spectral hue (blue points) and spectral hardness (pink points) for the 2007 full outburst. The vertical brown and gray lines and shaded region indicate the best-fitting values and 1-$\sigma$ contours for $t_{\rm hue, trans}$ and $t_{\rm hr, trans}$, respectively, and the shaded region indicates the uncertainty on the model fits. The four arrows indicate the four observations we model in sec.\ref{sec:results}. Right panel: identical plot for the 2009 hard state outburst. }
    \label{fig:hardness_hue}
\end{figure*}

For each outburst, we downloaded all available \RXTE data from the High Energy Astrophysics Science Archive Research Center (HEASARC) archive. The data was then reduced using the publicly available \texttt{Chromos} pipeline (\citealt{Gardenier18}, where the full details of the data reduction are provided). Briefly, \texttt{Chromos} extracts both spectra (in standard-2 modes) and lightcurves (in event, binned or good xenon modes, depending on the observation) from the Proportional Counter Array (PCA) following the standard \RXTE cookbook through a Python interface. Spectra are extracted from every layer of the second Proportional Counting Unit (PCU), as it is the best calibrated on the PCA \citep{Jahoda06,Shaposhnikov12,pcacorr}. Lightcurves are extracted between 3 and 13 keV, after which \texttt{Chromos} computes the power spectrum (correcting for Poisson noise and dead time) in segments of 512\,s each. A power spectrum for each observation is then computed by averaging over all segments, after which the pipeline computes the power colors and power spectral hue identically to \cite{Heil15a}. These authors define four continuous, logarithmically-spaced bands in Fourier frequency with boundaries at $0.0039$, $0.031$, $0.25$, $2.0$ and $16.0$ Hz, and then measure the variance in each by integrating the power spectrum. From the variances in each band, they then define two power colors as PC1 = variance (0.25--2.0 Hz)/variance (0.0039--0.31 Hz) and PC2 = variance (0.031--0.25)/variance (2.0--16.0 Hz). During an outburst, sources move in a loop around the $\log$(PC1)/$\log$(PC2) plane as the shape of the power spectrum evolves; as a result, one can use the location of each observation on this loop to characterize the shape of the power spectrum independently of its normalization. In practice, this is done by defining a center point (we use [4.51920, 0.453724] in linear units, identically to \citealt{Heil15a}), and computing the angle between a reference direction, which we (and \citealt{Heil15a}) take to be at a -135$^{\circ}$ angle starting from the positive $x$ axis, and each (PC1,PC2) point. This angle is defined as the power spectral hue. An example of where a typical HS (scarlet diamond) or HIMS observation would lie in the power color diagram is shown in Fig.\ref{fig:hueexample}, using the definitions above. If the power spectrum is flat (as in the low HS or SS) then PC1$\approx$PC2$\approx1$, corresponding the top left of the PC1-PC2 diagram, and the observation will have a hue near $0^{\circ}$. On the other hand, if the PSD is band-limited around a few Hz (as in the HIMS) then PC1$\gg1$ and PC2$\ll1$ and the points move to the bottom right of the PC1-PC2 plot, resulting in a hue of $\approx 180^{\circ}$.  

In each observation we then computed the source count rates (after background subtraction) between 3--6, 6--13 and 3--30 keV, the hardness ratio (taking 3--6 and 6--13 keV for the soft and hard band, respectively). Given that we are interested in whether each outburst results in a state transition, here we only present results for observations in the rise of the hard or intermediate state. We define these periods as any observation from the start of the observations, until the source reaches a hardness $<0.4$.

We selected four observations (ObsIDs 92052-07-03-01, 92052-07-06-01, 92035-01-02-03, and 92035-01-03-00) during the 2007 outburst of GX 339$-$4 to analyze in further detail, as a representative sample of observations during the rise. These are 3, 58, 75, and 79 days after the start of the X-ray observations, respectively, and were chosen to highlight the behavior of the source before and after the transitions in the power spectral hue and spectral hardness (described in the next section). We will refer to these observations as low/hard state (LHS, $t=3.\rm{d}$), bright hard state (BHS, $t=58.\rm{d}$), first intermediate state (HIMS1, $t=75.\rm{d}$) and second intermediate state (HIMS2, $t=79.\rm{d}$). For these observations, we re-binned the power spectra geometrically with a binning factor of $f=0.1$, in order to quantify in more detail their evolution over a broad range of Fourier frequencies. We corrected the time-averaged energy spectra for these epochs using the \texttt{PCACorr} tool and included an additional $0.1\%$ systematic error in order to reduce the remaining systematic features of the PCA as is commonly done \citep{pcacorr}. Both power spectra and time-averaged energy spectra were modeled in \texttt{Xspec}, version 12.11.1. 


\section{Results}\label{sec:results}

Here we present results of comparisons between the time-variability and spectral characteristics over time. The evolution of the power spectral hue (blue points) and spectral hardness (pink points) for the 2007 (full) and 2009 (hard state) outbursts are shown in Fig.\ref{fig:hardness_hue}. In the former (left panel), the state transition is marked by an increase in power spectral hue and drop in spectral hardness, which is not present in the latter (right panel). However, the power spectral hue begins evolving roughly two weeks earlier than the spectral hardness, meaning that the state transition first affects the power-spectrum, with the time-averaged energy spectrum changing later. \textit{This behavior is repeated in every outburst of GX 339$-$4 we analyzed, as shown in the Appendix in Fig.\ref{fig:hardness_hue_1}}. 
\begin{figure*}
    \centering
    \includegraphics[width=\columnwidth, trim={0.65cm 0.0cm 0.6cm 0.0cm},clip]{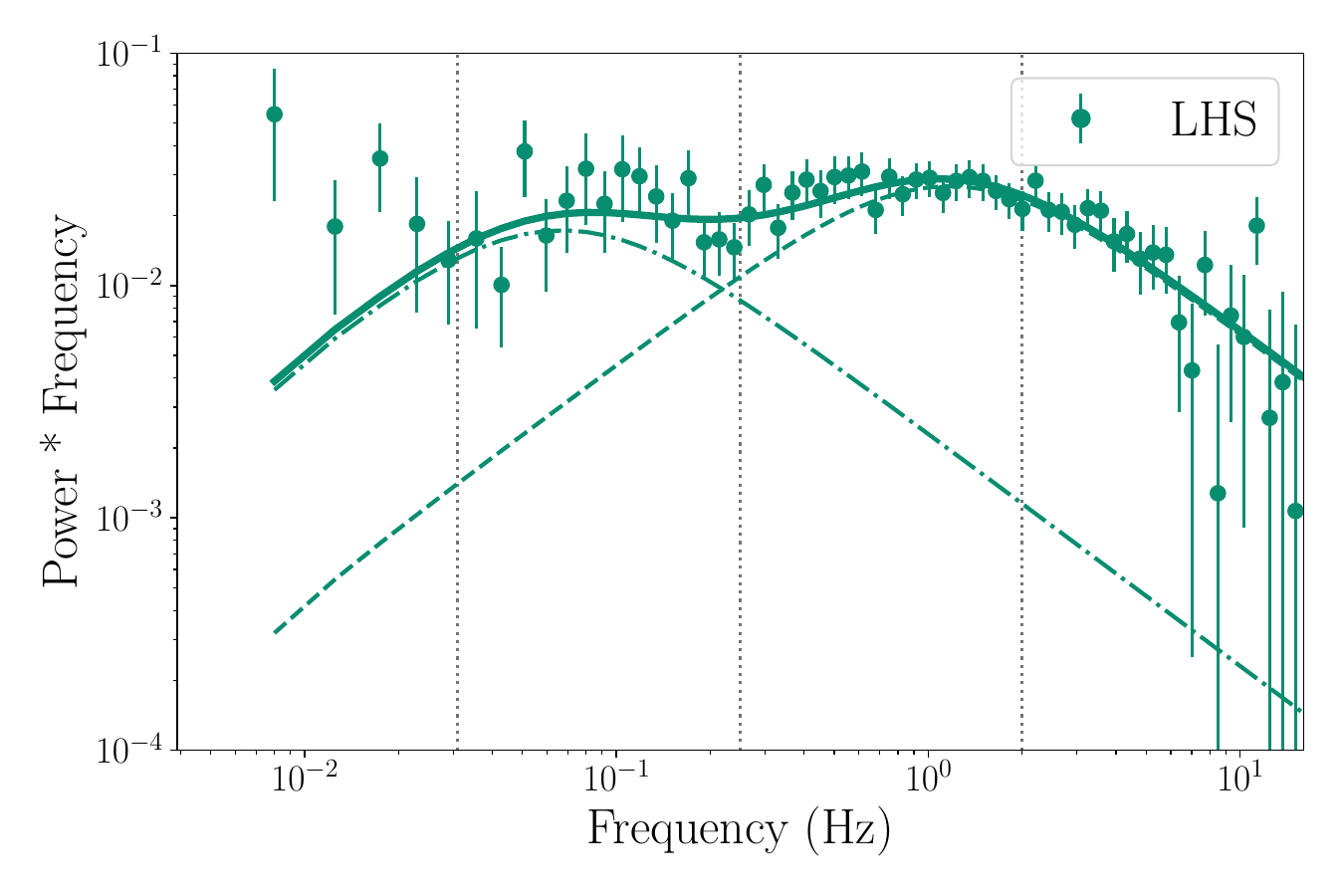}
    \includegraphics[width=\columnwidth, trim={0.65cm 0.0cm 0.6cm 0.0cm},clip]{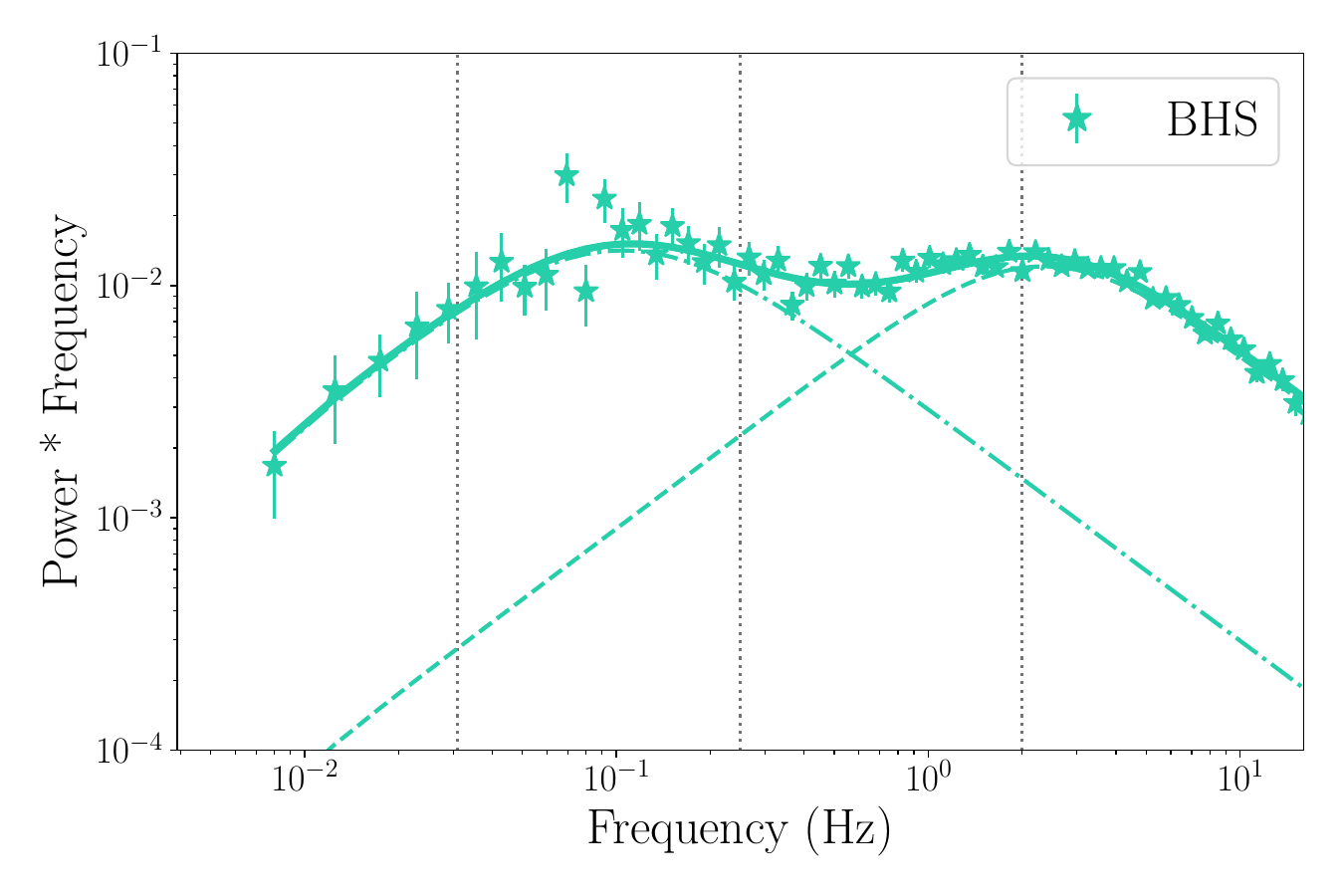}
    \includegraphics[width=\columnwidth, trim={0.65cm 0.0cm 0.6cm 0.0cm},clip]{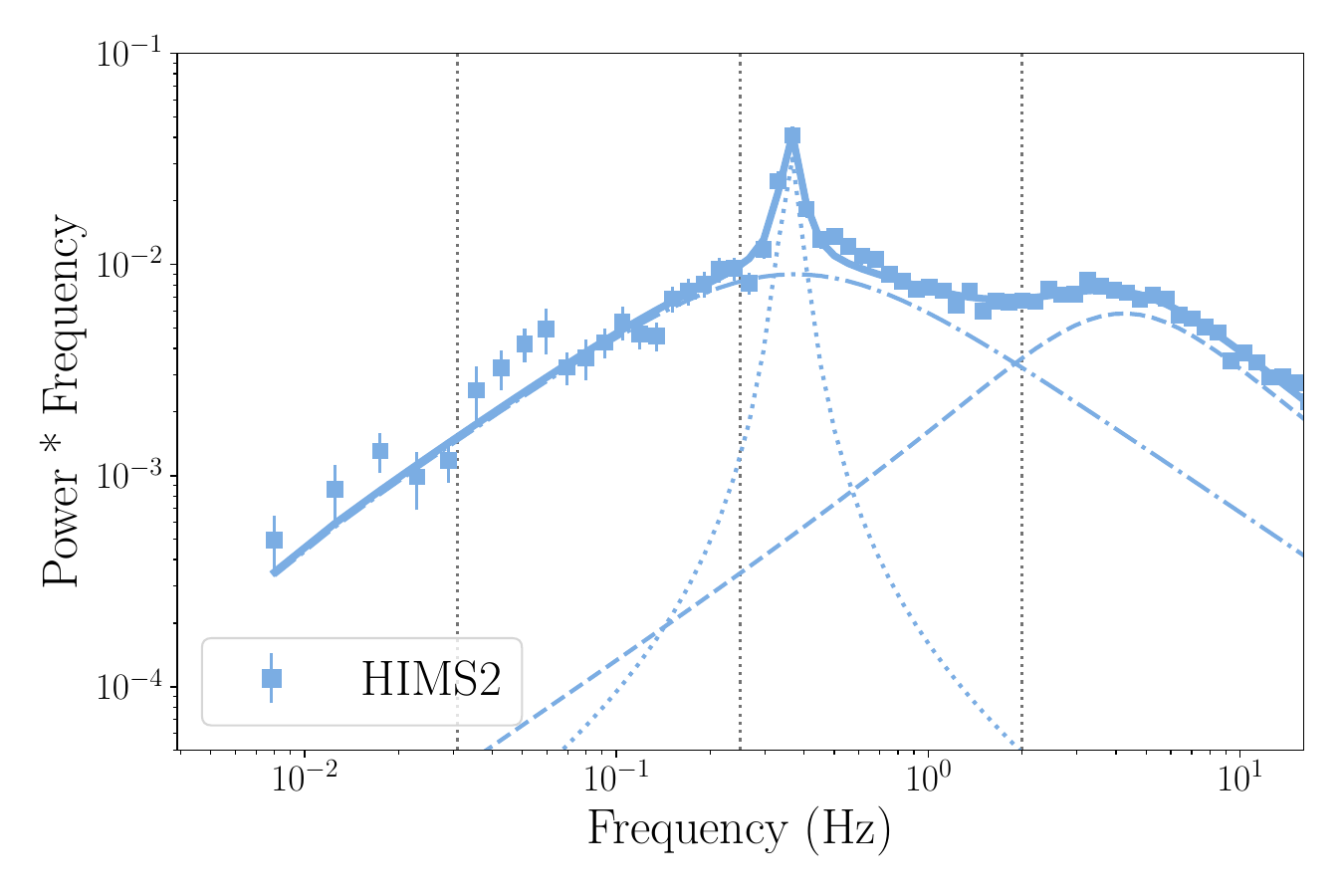}
    \includegraphics[width=\columnwidth, trim={0.65cm 0.0cm 0.6cm 0.0cm},clip]{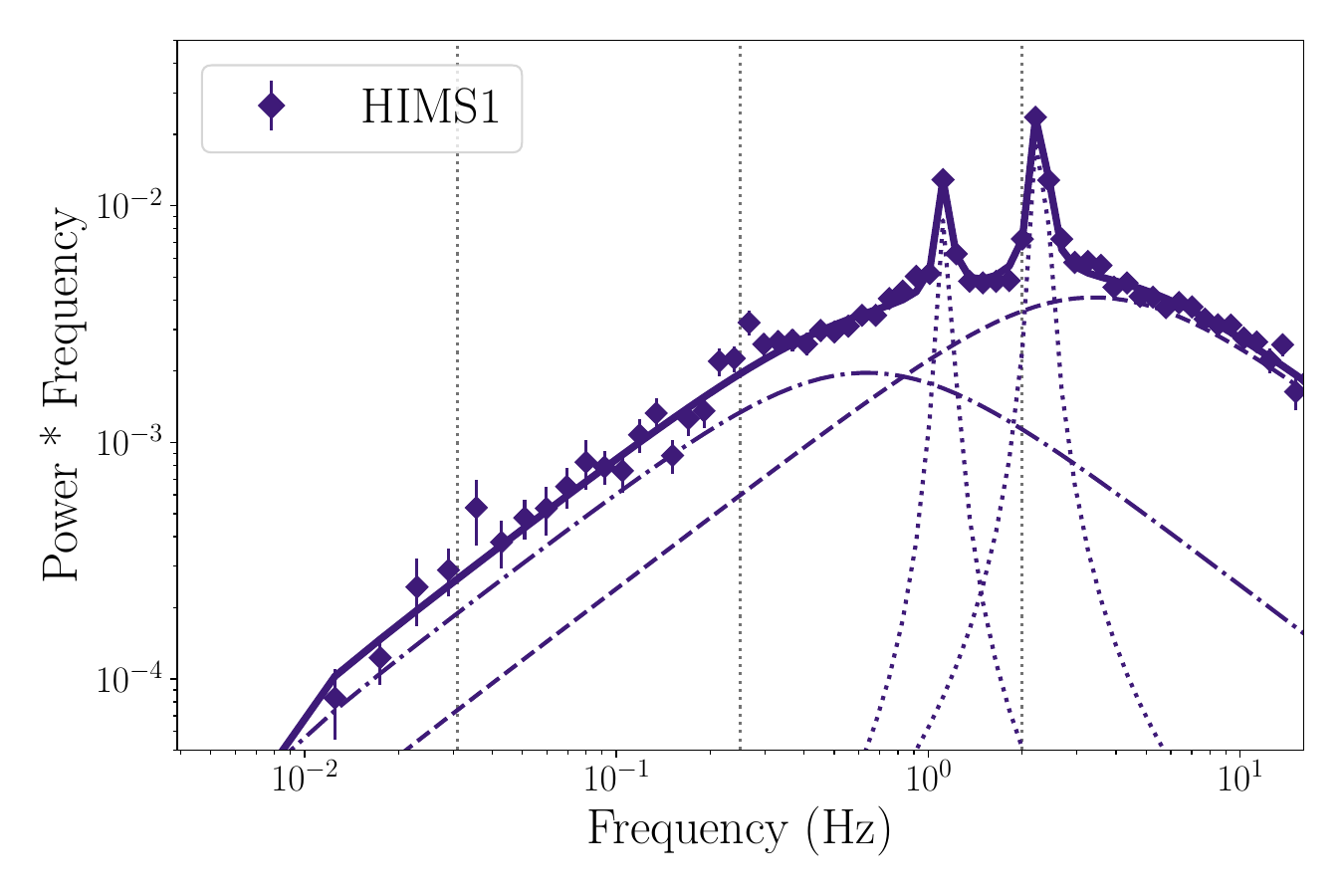}
    \caption{Power spectra of the four epochs highlighted in Fig.\ref{fig:hardness_hue}. The top left observation was taken at the start of the outburst, the top right immediately before $t_{\rm hue,trauns}$, the bottom left between the transition times $t_{\rm hue,trauns}$ and $t_{\rm hr,trauns}$, and the bottom right immediately after $t_{\rm hr,trauns}$. The vertical lines indicate the frequency bounds used to compute the power spectral hue. Dash-dotted lines indicate the zero-centered Lorentzian and dashed lines the broad Lorentzian used to model the broadband noise. Dotted lines indicate the narrow Lorentzians used to model the QPOs. The solid lines represents the total model.}
    \label{fig:powerspectra}
\end{figure*}
For the outbursts of GX 339$-$4 we quantified this behavior by fitting either linear (for hard state outbursts) or double linear (for full outbursts) models to the time evolution of both power spectral hue and spectral hardness, using the \texttt{scipy} function \texttt{curve${\_}$fit}. The double linear model is defined as:
\begin{equation}
\rm{hue}(t) = \left\{
\begin{aligned}
& m_{1}t + h_{\rm hue,0} \quad & t < t_{\rm hue,trans}  \\
& m_{2}(t-t_{\rm hue,trans}) + \\
& m_{1}t_{\rm hue,trans} + h_{\rm hue,0} \quad &  t \geq t_{\rm hue,trans}
\end{aligned}
\right. 
\end{equation}
where $m_{1}$ and $m_{2}$ are the slopes before and after the transition time $t_{\rm hue,trans}$, and $h_{hue,0}$ is the power spectral hue at $t=0$, which for these fits we define as the start of the first RXTE observation. We define the model parameters for the spectral hardness identically, using the subscript ${\rm hr}$ instead. The best-fitting values for $t_{\rm hue,trans}$ and $t_{\rm hr,trans}$ are reported in Tab.\ref{tab:outburst_date_1} in the Appendix. We note that for the 2002 outburst, the power spectral hue is already rising at the start of the observations, and therefore the constraints on $t_{\rm hue,trans}$ are far worse than in the other outbursts. Regardless, in all outbursts the spectral hardness transition time $t_{\rm hr,trans}$ is systematically larger than the power spectral hue transition time $t_{\rm hue,trans}$ by $\approx 10-40$ days.

The four observations from the 2007 outburst we selected are marked with arrows in Fig.\ref{fig:hardness_hue}. These were chosen so that one observation is at the beginning of the hard state (LHS, green), one is shortly before $t_{\rm hue,trans}$ (BHS, cyan), one is shortly before $t_{\rm hr,trans}$ (HIMS1, light blue), and one is immediately after $t_{\rm hr,trans}$ (HIMS2, purple). We fitted the power spectrum from each with a broad zero-centered Lorentzian to account for the low-frequency noise, one additional broad Lorentzian for the high-frequency noise, and added additional narrow Lorentzians for the quasi periodic oscillations (QPOs) as necessary. These fits are shown in Fig.\ref{fig:powerspectra}, and the best-fitting parameters are reported in the Appendix in Tab.\ref{tab:PSD_fits}. 

There is a noticeable evolution in the low-frequency broadband noise, which shifted to increasingly high frequencies at similar root mean square (rms) roughly until $t_{\rm hr,trans}=77\,\rm{d}$ (first three epochs), before decreasing in rms at a similar frequency; similar behavior has been reported previously in various sources \citep[e.g.][]{Cui99,Pottschmidt03,Belloni05}. At the same time, the high frequency noise shifted more subtly towards higher frequencies, and its rms slowly decreased. Type-C QPOs appeared around $t_{\rm hr,trans}$, but these are not expected to dramatically affect the power spectral hue \citep{Heil15b}. Thus, the change in power spectral hue we observe is driven mainly by the decrease in long-timescale variability (quantified by the low frequency Lorentzian). 
\begin{figure}
    \centering
    \includegraphics[width=\columnwidth, trim={0.65cm 0.0cm 0.6cm 0.0cm},clip]{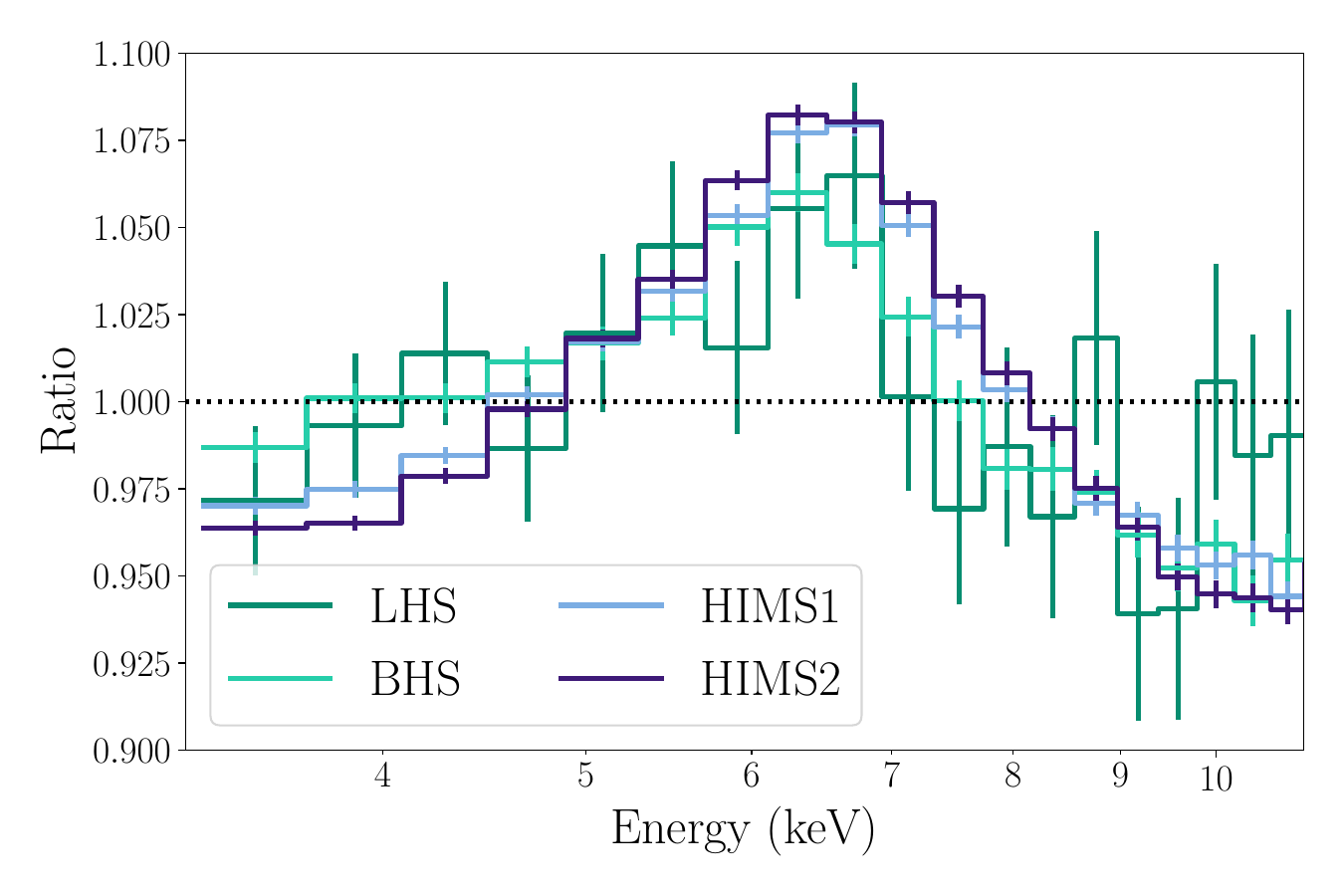}
    \caption{The iron line profile in the four epochs highlighted in Fig.\ref{fig:hardness_hue}, shown as the ratio to a Comptonization model. The color coding is identical to Fig.\ref{fig:powerspectra}. We only show the 3-11 keV band for clarity.}
    \label{fig:Ironline}
\end{figure}

The time-averaged energy spectra, on the other hand, vary more subtly. Fig.\ref{fig:Ironline} shows the ratio plots for each spectrum fitted (in the 3-30 keV band) with a simple absorbed Comptonization model (\texttt{tbabs*nthcomp} in Xspec). The iron profile remains similarly broad in every epoch. We also jointly fitted these spectra with a reflection model (\texttt{tbabs*(nthcomp+relxillCp)} in Xspec); we assume a maximal spin of $a{*}=0.998$ and tied the inclination and Iron abundance among all epochs. The fits are relatively insensitive to the column density $nH$, which we froze to $nH=0.7$ \citep[e.g.][]{Garcia15} in all epochs. Finally, the fit to the LHS spectrum is insensitive to the emissivity profile $q$, which we froze to $q=3$. The best-fitting parameters are reported in Tab.\ref{tab:spec_fits} and the fits are shown in the Appendix in fig\ref{fig:energyspectra}. We find consistently small truncation radii for the disk, as is common with reflection modeling \citep[e.g.][]{Garcia15,Connors19,Sridhar20,Wang21}. Therefore, it appears that while the X-ray variability was changing significantly during these observations, the innermost accretion flow was not. 

\section{Discussion and conclusions}\label{sec:discussion}

We have shown that during full BHXRB outbursts, the low-frequency variability begins dropping $\approx10-40$ days ahead of any spectral change. This behavior is never displayed by hard state outbursts. In other words, \textit{one can tell whether the source will transition or not from the properties of the X-ray variability, before any large change in the hardness ratio or time-averaged energy spectrum.} This is the main result of our paper. Our results are in contrast to those of \cite{Alabarta21}, who found X-ray data is not sufficient to predict the state transition, purely due to how we visualized our data - to our knowledge, our work is the first to compare the time evolution of the power spectral hue and spectral hardness.

In particular, we find that the main change leading to the state transition is a decrease in the variability at low Fourier frequency. In the standard model of propagating fluctuations \citep[e.g.][]{Lyubarskii97,Arevalo06,Ingram11,Rapisarda16}, a decrease in low frequency variability indicates that the outer regions of the accretion flow generated less variability and become progressively more stable. Therefore, our findings suggest that the state transition is driven by a decrease in the turbulence generated from the outer thin disk, which is effectively divided in two parts: an inner, turbulent flow responsible for generating the low frequency variability, and an outer, stable flow. The transition from stable to turbulent flow can be thought of as a ``quiet front''; as the luminosity increases this ``quiet front'' propagates inwards, causing the accretion flow to stabilize progressively until the soft state is reached. Indeed, it is well established that the direct thin disk emission is highly variable in the hard state \citep{WIlkinson09,Uttley11} and remarkably stable in the soft state \citep{Belloni05}; our findings indicate that the transition between the two types of thin disks is likely linked to the state transition. 

Following \cite{Churazov01,Done07,Ingram11}, we can estimate the radius $R_{\rm trans}$ at which the thin disk is switching from turbulent to stable, by associating the peak of the low-frequency Lorentzian $\nu_{\rm l}$ to the viscous frequency (the reciprocal of the viscous timescale) $\nu_{\rm visc} = \alpha (h/r)^2 \nu_\phi$, where $h/r$ is the (thin) disk aspect ratio, $\alpha$ the disk viscosity and $\nu_\phi$ is the (Keplerian) orbital frequency. Using the relativistic expression for the orbital frequency and re-arranging gives: 
\begin{equation}
R_{\rm trans} = \left[\frac{(h/r)^{2}\alpha c - 2\pi R_{\rm g}a^{*}\nu_{\rm l}}{2\pi R_{\rm g}\nu_{\rm l}}\right]^{2/3}\,R_{\rm g},    
\end{equation}
where $c$ is the speed of light and $a^{*}$ the black hole spin. From the fits to the power spectra in Fig.\ref{fig:powerspectra} we can estimate $\nu_{\rm l}\approx 0.06$, $0.1$, $0.36$ and $0.5$ Hz, respectively. Assuming an $8\,M_{\odot}$ black hole and conservatively\footnote{Smaller values of $h/r$ and $\alpha$ and larger spins both result in smaller inferred values for $R_{\rm trans}$.} taking $h/r=\alpha=0.1$ and $a^{*}=0$, we find $R_{\rm trans}\approx 16$, $11$, $6$ and $4\,$ $R_{\rm g}$ respectively for each observation. While we note that in the first epoch there are hints of additional noise at frequencies below the first Lorentzian, meaning $R_{\rm trans}$ could be larger, these in general constitute upper limits on $R_{\rm trans}$, as larger values of $h/r$ or $a^{*}$ will result in lower radius estimates. These values of $R_{\rm trans}$ are larger than, but on the order of, the estimates for $R_{\rm in}$ we found from modeling the time-averaged energy spectra (Tab.\ref{tab:spec_fits}), indicating that both X-ray variability and reflection originate close to the black hole.

Alternatively, one can associate the (much longer) timescale for the state transition $t_{\rm hr,trans}-t_{\rm hue,trans}$ with the viscous time scale \citep{Frank02,Done07}, which we can re-write in the $a^{*}=0$ limit as:
\begin{equation}
t_{\rm visc}=4.5\alpha^{-1}\left(\frac{h}{r}\right)^{-2}\frac{m}{10}\left(\frac{R_{\rm trans}}{6}\right)^{3/2}\,\rm{ms}, 
\label{eq:t_visc}
\end{equation}
where $m$ is the mass of the black hole in units of Solar masses and $r$ the radius in units of $R_{\rm g}$. Setting $t_{\rm visc}=t_{\rm hr,trans}-t_{\rm hue,trans}\approx10-40\,\rm{days}$ and solving for the radius (again assuming $h/r=\alpha=0.1$ and an $8\,M_{\odot}$ black hole) gives $R_{\rm trans}\approx 10^{4-5}\,\rm{R_g}$, a far larger radius. This is not surprising: as eq.\ref{eq:t_visc} shows, long times inevitably correspond to large radii. In other words, if the state transition is caused by a viscous process over timescales of weeks, then this process must originate from the outermost regions of the disk. Interestingly, \cite{Homan05b} reported a similar $\approx2$ week delay between the optical/IR and X-ray emission (which originate in the outer and inner disk, respectively) in the SS (after the state transition has occurred) during the 2002 outburst. 

The presence in the hard state of a turbulent disk, strong corona and powerful jet all point to the presence of dynamically important magnetic fields in the accretion flow during the HS \citep[e.g.][]{Liska20,Dexter21,Liska22}; vice versa, the lack of these observational features in the SS suggests that in this state the flow is not strongly magnetized. We can then speculate that the flow in the HS is in the Magnetically Arrested Disk (MAD) regime \citep{Narayan03}, and that the ``quiet front'' driving the state transition is a relatively stable, un-magnetized region of the disk that is slowly moving towards the black hole. Eventually, as the source reaches the SS, the entire flow leaves the MAD state and the corona and jet are severely weakened or entirely quenched. This picture is similar to that proposed by \cite{Begelman14}. 

\begin{figure}
    \centering
    \includegraphics[width=\columnwidth, trim={0.65cm 0.0cm 0.6cm 0.0cm},clip]{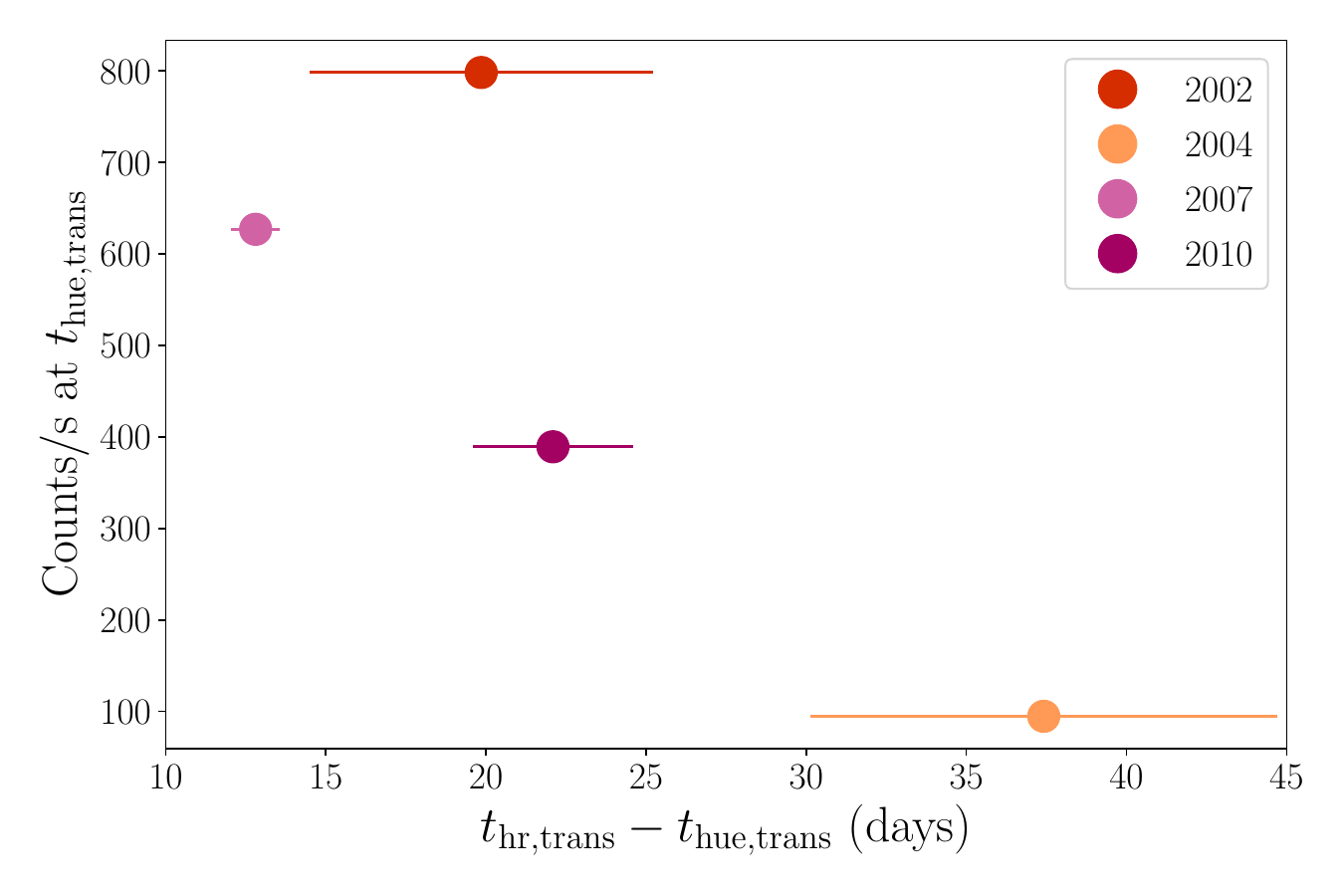}
    \includegraphics[width=\columnwidth, trim={0.65cm 0.0cm 0.6cm 0.0cm},clip]{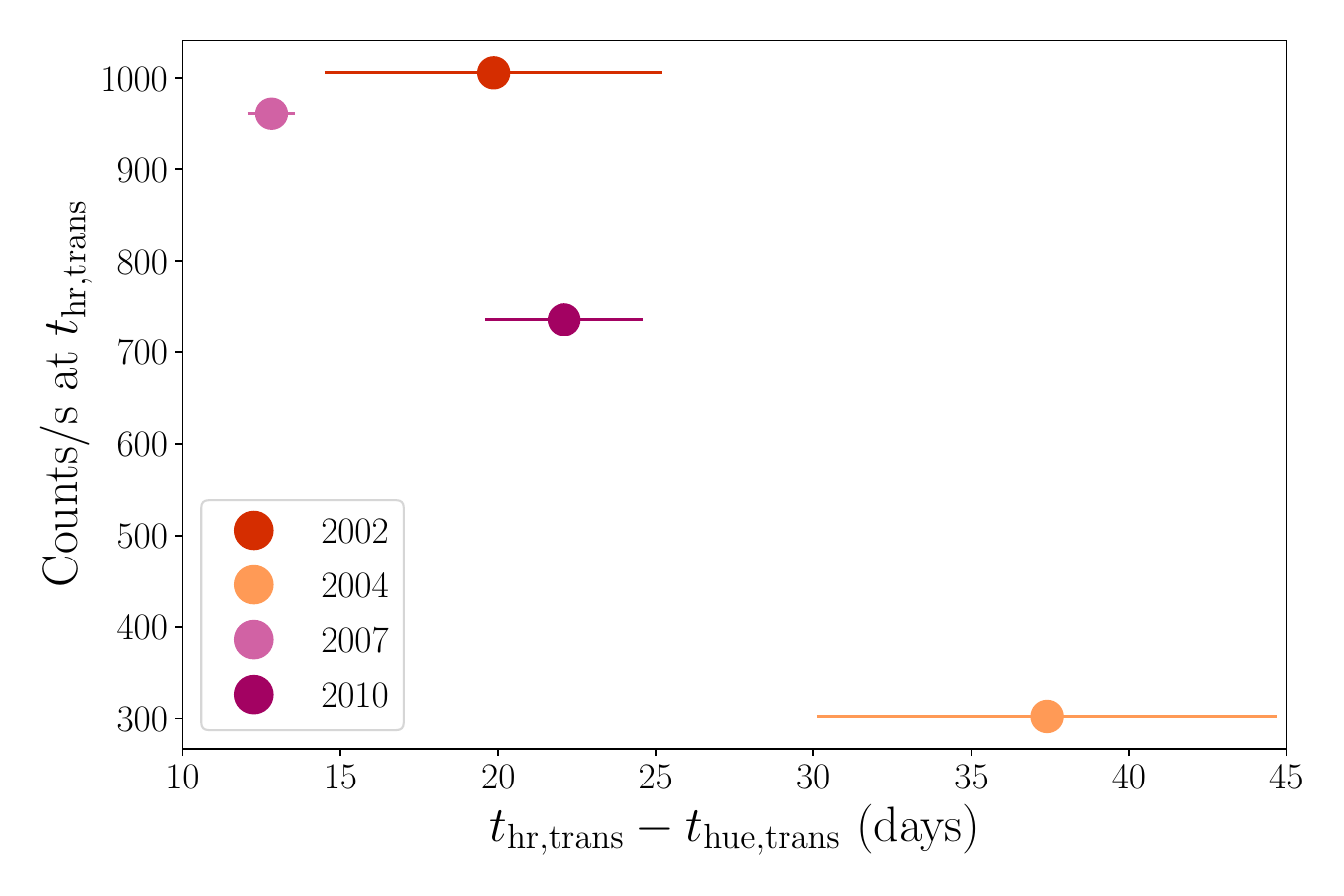}
    \caption{Difference in transition times for each full outburst of GX 339$-$4 against the 3-30 keV count rate at the transition times $t_{\rm hue,trans}$ (top panel) and $t_{\rm hr,trans}$ (bottom panel).  The error bar for the count rate is smaller than the symbols. Brighter outbursts tend to show quicker state transitions.}
    \label{fig:lagvsrate}
\end{figure}

The outbursts of GX 339$-$4 are notable because of their recurring nature, the excellent \RXTE coverage during both the HS rise and state transition, and because the latter does not always happen at the same luminosity \citep[e.g.][]{Alabarta21}. Comparing all the full outbursts observed by \RXTE, we find that outbursts transition faster if the source switches states at higher luminosities. This is shown in Fig.\ref{fig:lagvsrate}, in which we plot the difference in transition times $t_{\rm hr,trans} - t_{\rm hue,trans}$ against the source count rate when the state transition occurs, \footnote{For the 2002 outburst, $t_{\rm hr,trans} - t_{\rm hue,trans}$ is poorly constrained due to the lack of observations during the rise.} defined either as the count rate at $t_{\rm hue,trans}$, or that at $t_{\rm hr,trans}$. While the sample size is limited, the two appear to be anti-correlated. Assuming that the radiative efficiency is unchanged between outbursts, this finding implies that whatever instability drives the system evolution operates faster at higher accretion rates. This trend is similar to that found by \cite{Yu09}, who found that the luminosity during the state transition is correlated with the rate of increase in the X-ray luminosity during the rising hard state.

Finally, we note that the trend of evolving power spectral  hue ahead of the canonical state transition appears to be applicable to BHXRBs as a whole, for two reasons. First, the evolution of the additional sources we analyzed is shown in the Appendix in Fig.\ref{fig:hardness_hue_all_1} and Fig.\ref{fig:hardness_hue_all_2}. These sources display behavior consistent with GX 339$-$4: if the power spectral hue is large at the start of the available observations, then a state transition inevitably occurs (e.g. XTE J1550$-$564 in 1999 and 2000, panels a and b of Fig.\ref{fig:hardness_hue_all_2}; H1743$-$322, panels c, d and e of Fig.\ref{fig:hardness_hue_all_1}). If instead the power spectral hue early on is low, the state transition only occurs after it has reached large values (e.g. GRO 1655-40, XTE J1752-223, panels a and b of Fig.\ref{fig:hardness_hue_all_1} although in the latter case observations are extremely sparse). If the hue does not evolve, then state transitions never occur (XTE J1550$-$564 in 2001, 2002 and 2003, panels c, d and e of Fig.\ref{fig:hardness_hue_all_2}). Unfortunately, the more sparse observational coverage for these additional sources makes it impossible to estimate $t_{\rm hue,trans}$. The outburst that is most similar to GX 339$-$4 is that of GRO 1655$-$40, which shows the same rising trend in the power spectral hue while the spectral hardness remains constant. Second, Fig.2 of \cite{Heil15b} (who analyzed an even larger sample of sources) shows that for a near-constant hardness of $\approx 1$ sources display a wide range of power spectral hues, and demonstrated that this trend is not dependent upon inclination. While lacking the time evolution of both quantities, this plot indicates that the power spectral hue consistently varies at near-constant spectral hardness, ahead of the canonical state transitions.

To conclude, we have demonstrated for the first time that it is possible to predict in a model-independent fashion whether a source will leave the hard state, before any spectral changes occur. These findings indicate that the state transition is driven by a decrease in the turbulence in the outer disk, and that this change occurs more quickly at higher accretion rates. Our results will also enable observers to more effectively coordinate observational campaigns of these systems, providing further insight on the evolution of their accretion flows. 

\begin{acknowledgments}
\section*{Acknowledgments}
We are thankful to the anonymous reviewer, whose comments have improved the quality and clarity of the manuscript. ML, GM, JW, EK and JAG acknowledge support from NASA~ADAP grant 80NSSC17K0515. JW acknowledges support from the NASA~FINNEST Graduate Fellowship, under grant 80NSSC22K1596. AI acknowledges support from the Royal Society. MK acknowledges support by the NWO Spinoza Prize. OK acknowledges funding by the Deutsches Zentrum f{\"u}r Luft-und Raumfahrt, contract 50 QR 2103. This research has made use of data and software provided by the High Energy Astrophysics Science Archive Research Center (HEASARC), which is a service of the Astrophysics Science Division at NASA/GSFC. 
\end{acknowledgments}

\vspace{5mm}


\software{Heasoft \citep{Heasoft}), XSPEC \citep{Arnaud96}, Chromos \citep{Gardenier18}, Stingray (\citealt{Stingray1}, \citealt{Stingray2}), PCACorr \cite{pcacorr}, Numpy \citep{numpy}, Matplotlib \citep{matplotlib}, Scipy \citep{scipy}.}



\appendix
\renewcommand\thefigure{\thesection.\arabic{figure}}  
\renewcommand\thetable{\thesection.\arabic{table}}  
\setcounter{figure}{0}   
\setcounter{table}{0}   

\section{Appendix A}

\begin{figure*}[h]
    \centering
    \includegraphics[width=0.49\columnwidth, trim={0.65cm 0.0cm 0.6cm 0.0cm},clip]{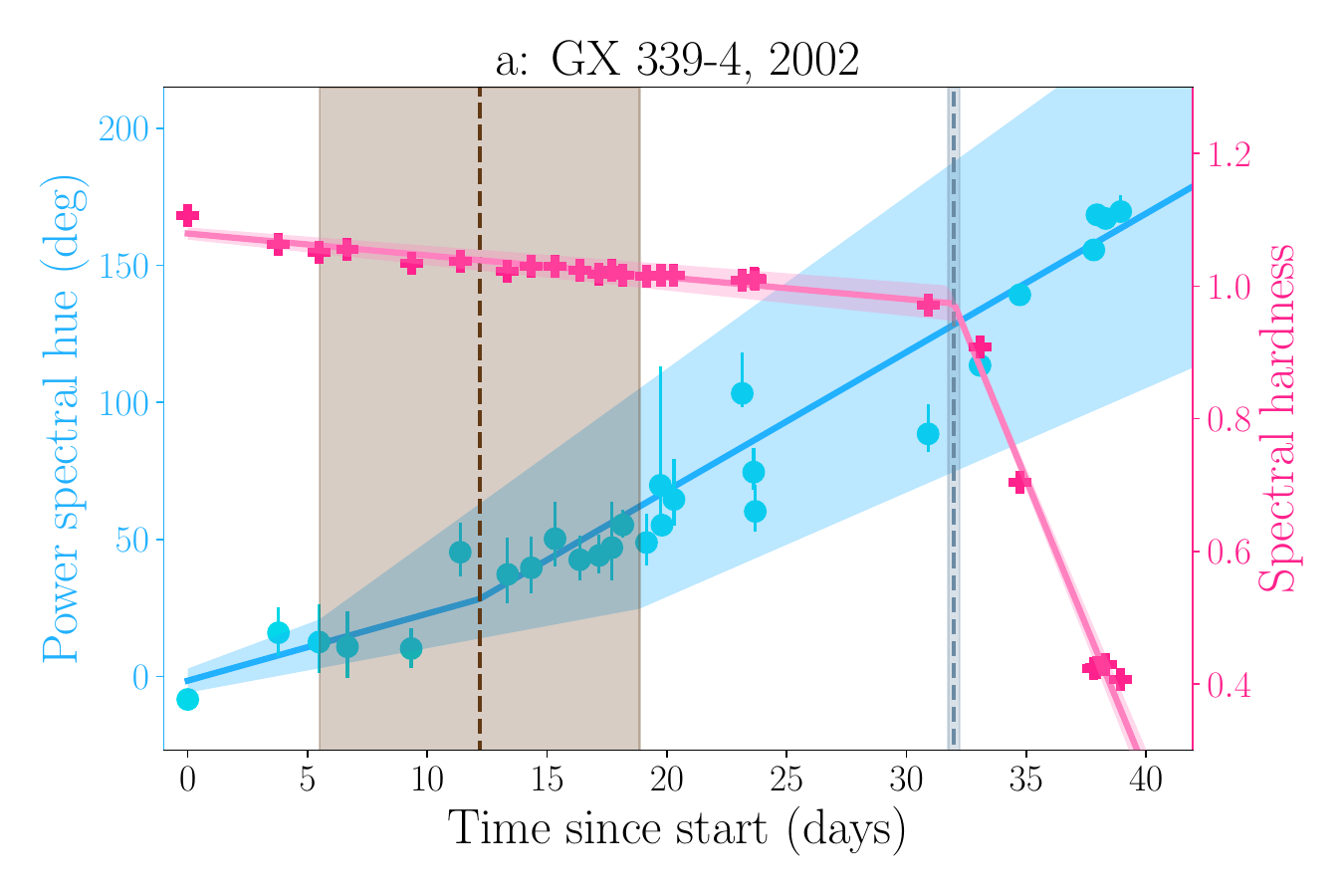}
    \includegraphics[width=0.49\columnwidth, trim={0.65cm 0.0cm 0.6cm 0.0cm},clip]{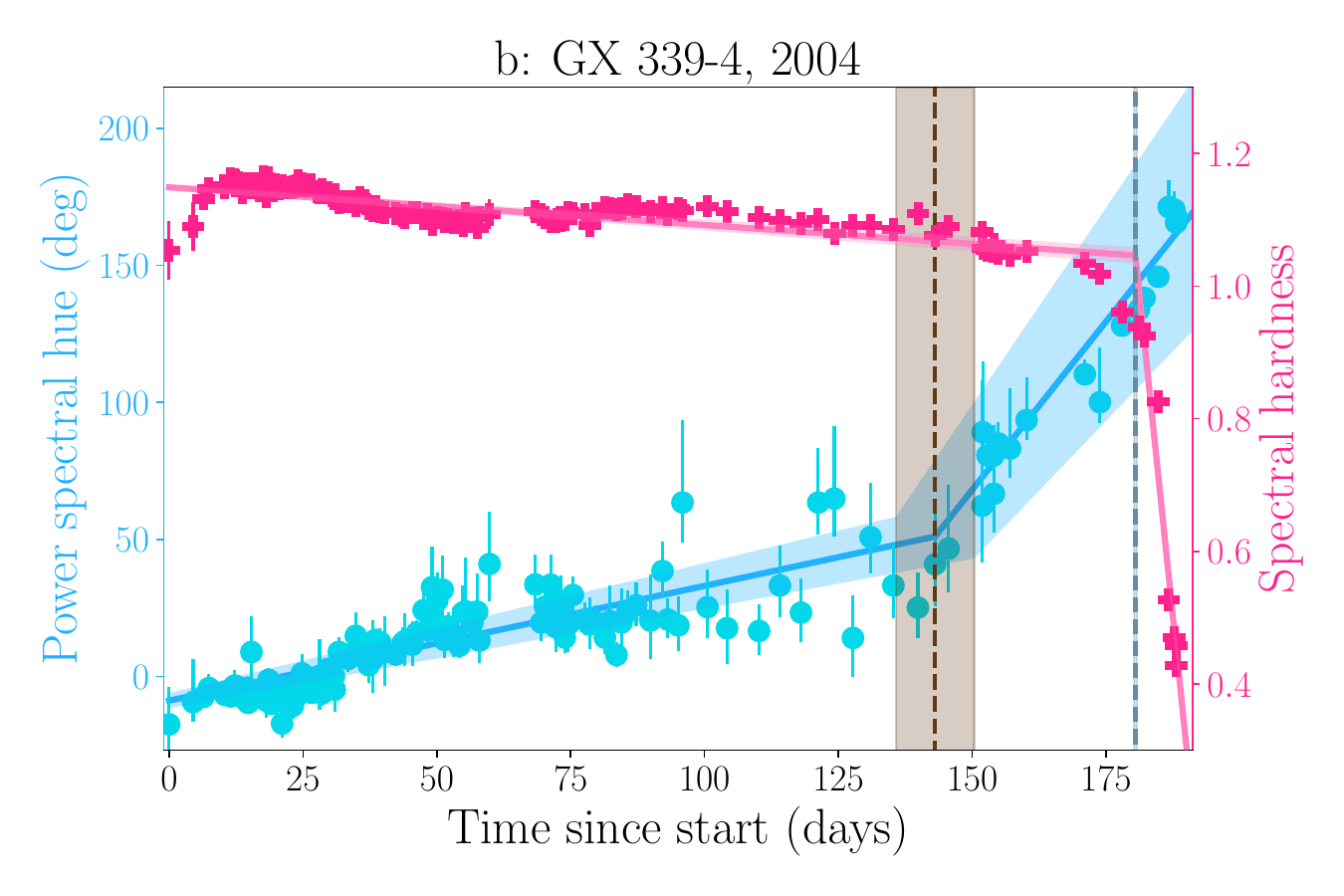}
    \includegraphics[width=0.49\columnwidth, trim={0.65cm 0.0cm 0.6cm 0.0cm},clip]{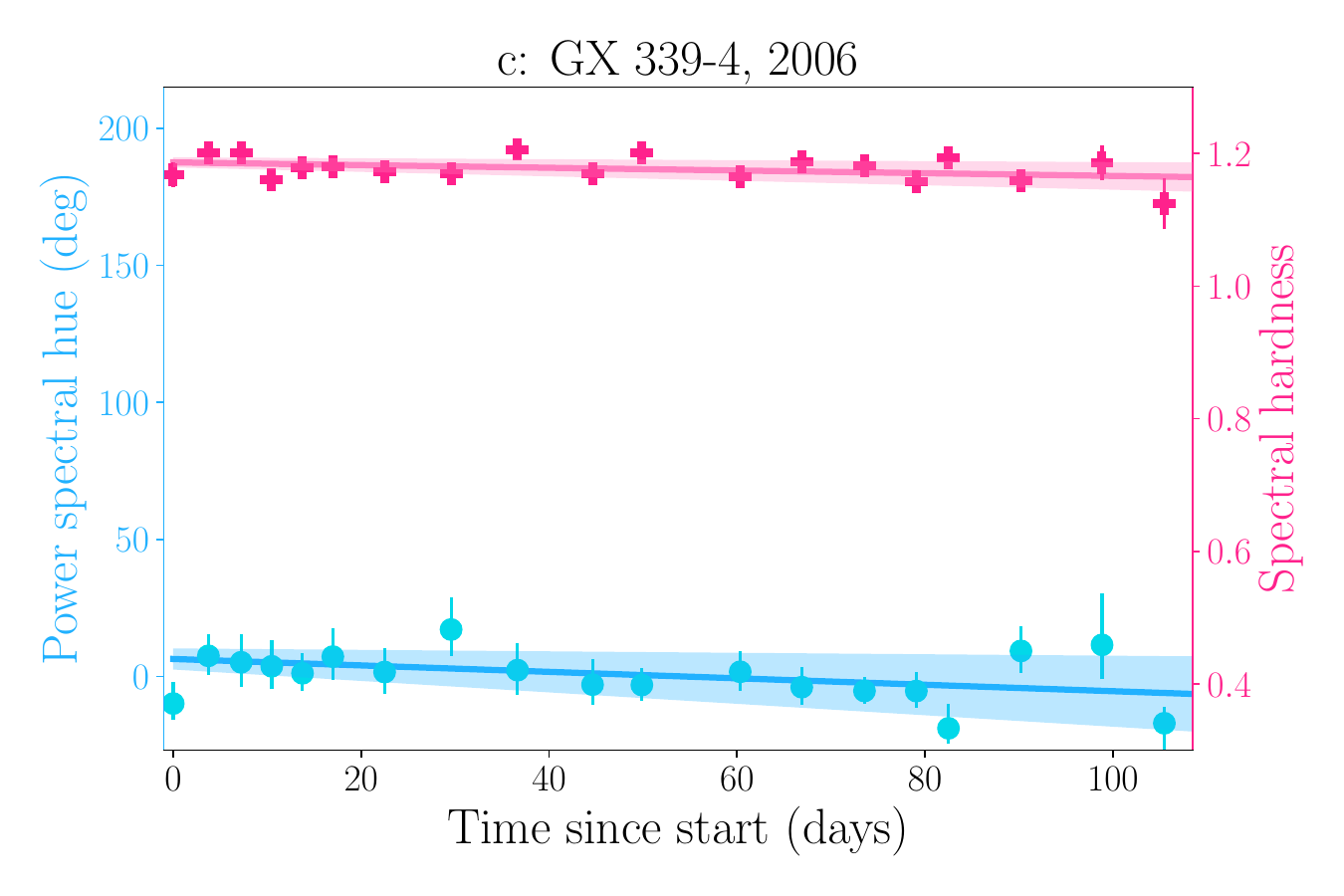}
    \includegraphics[width=0.49\columnwidth, trim={0.65cm 0.0cm 0.6cm 0.0cm},clip]{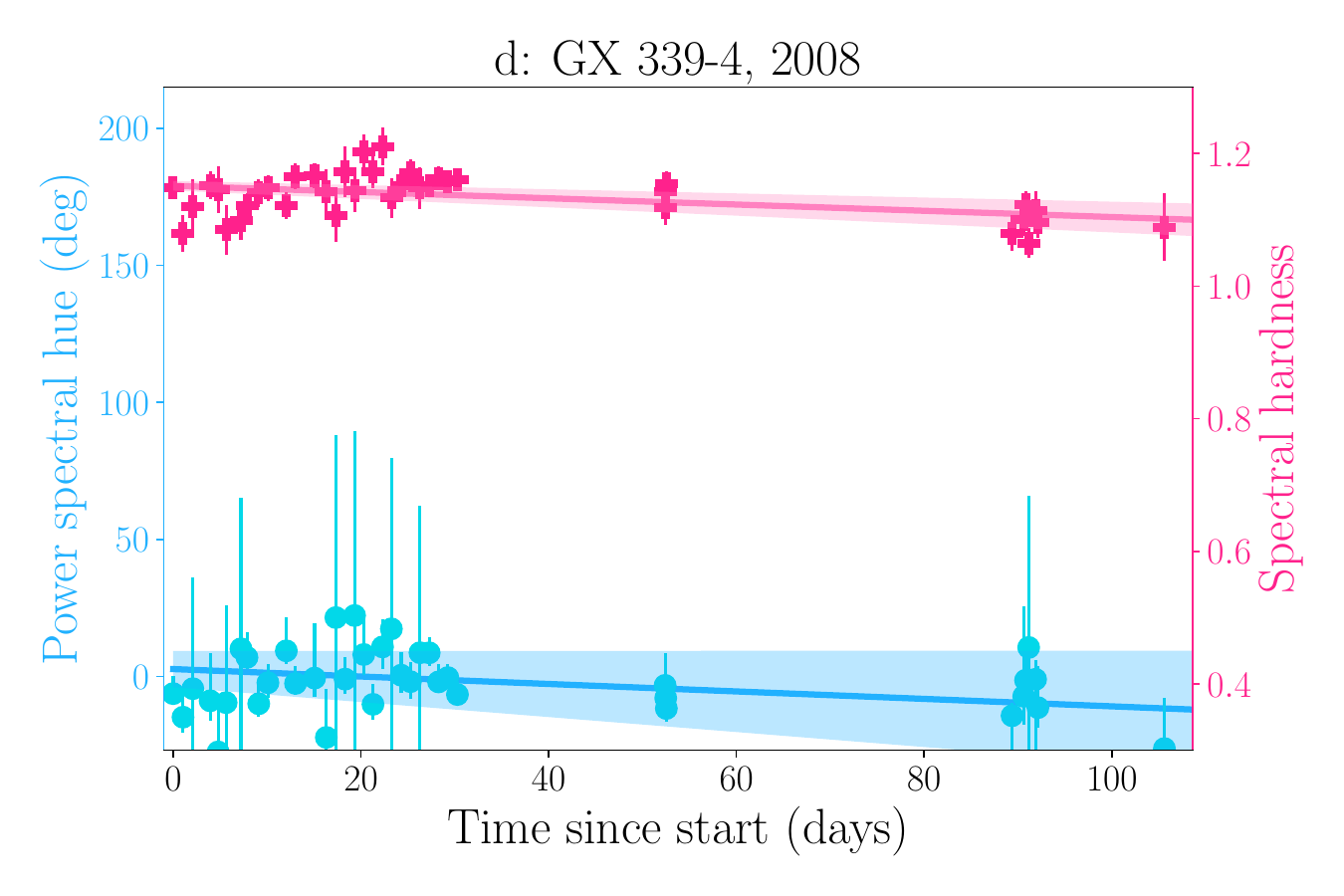}
    \includegraphics[width=0.49\columnwidth, trim={0.65cm 0.0cm 0.6cm 0.0cm},clip]{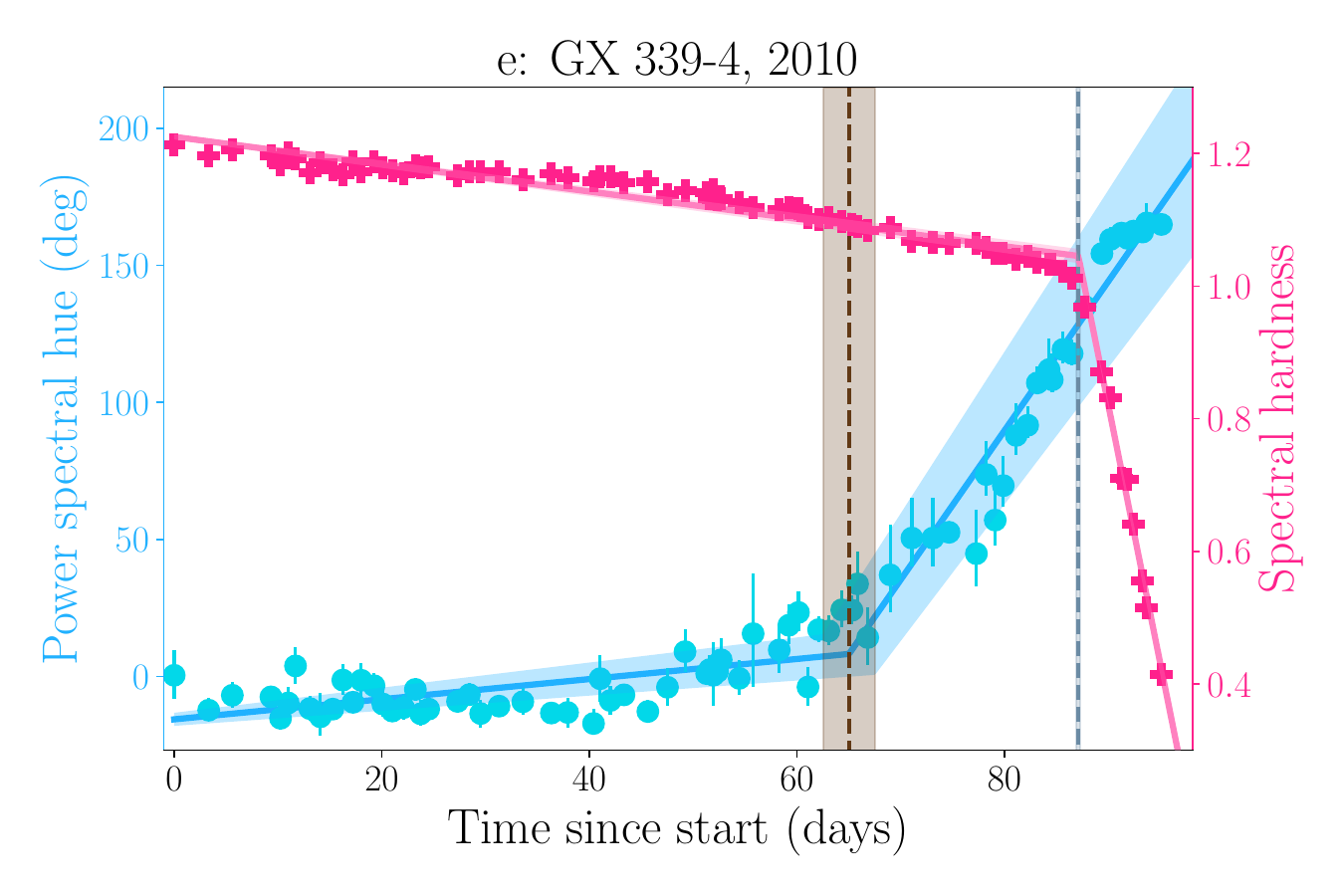}
    \caption{Evolution of the power spectral hue and spectral hardness  for the 2002, 2004, 2006, 2008 and 2010 outbursts of GX-339. The color coding is identical to Fig.\ref{fig:hardness_hue}.}
    \label{fig:hardness_hue_1}
\end{figure*}

\begin{figure*}[h]
    \centering
    \includegraphics[width=0.49\columnwidth, trim={0.65cm 0.0cm 0.6cm 0.0cm},clip]{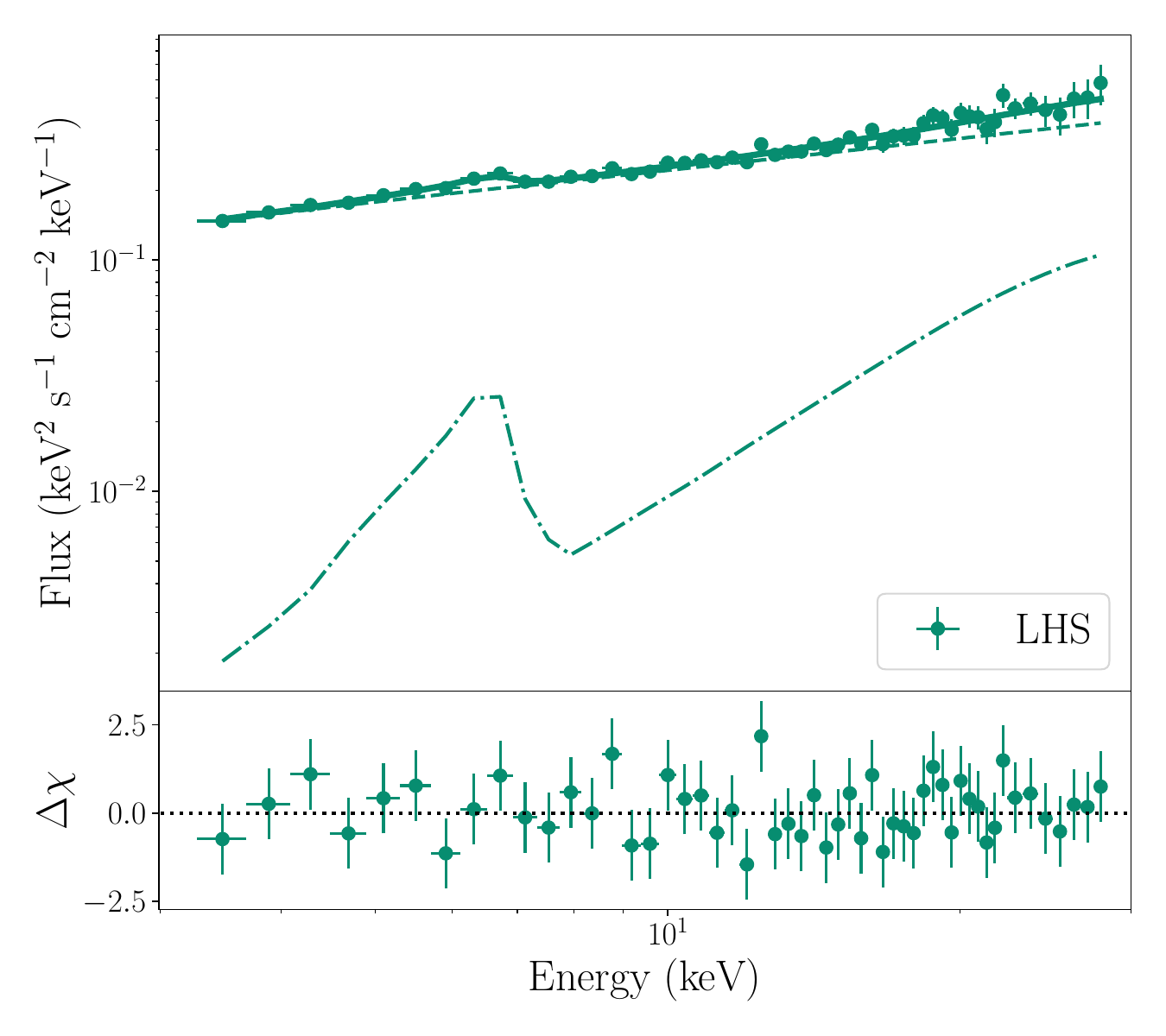}
    \includegraphics[width=0.49\columnwidth, trim={0.65cm 0.0cm 0.6cm 0.0cm},clip]{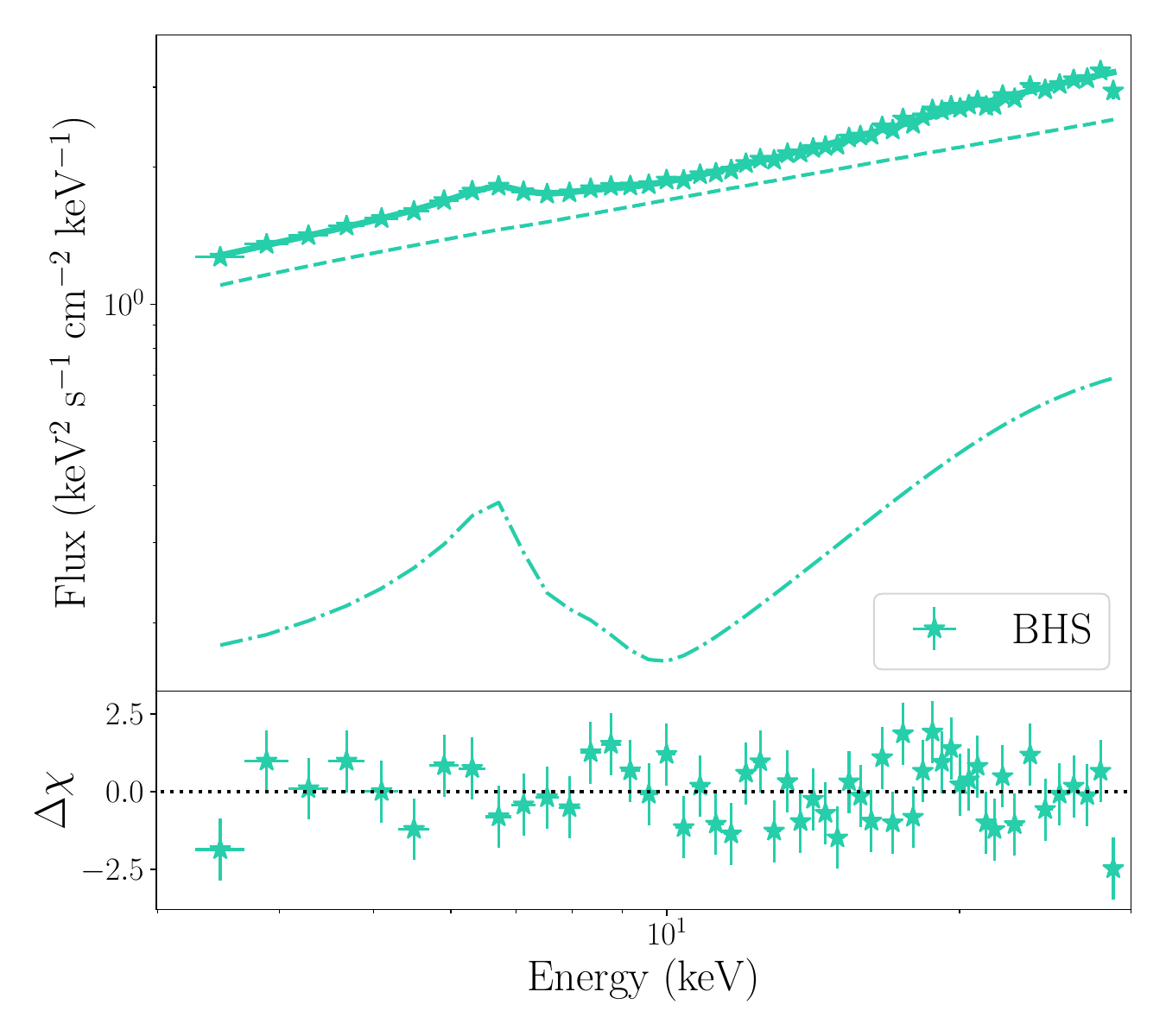}
    \includegraphics[width=0.49\columnwidth, trim={0.65cm 0.0cm 0.6cm 0.0cm},clip]{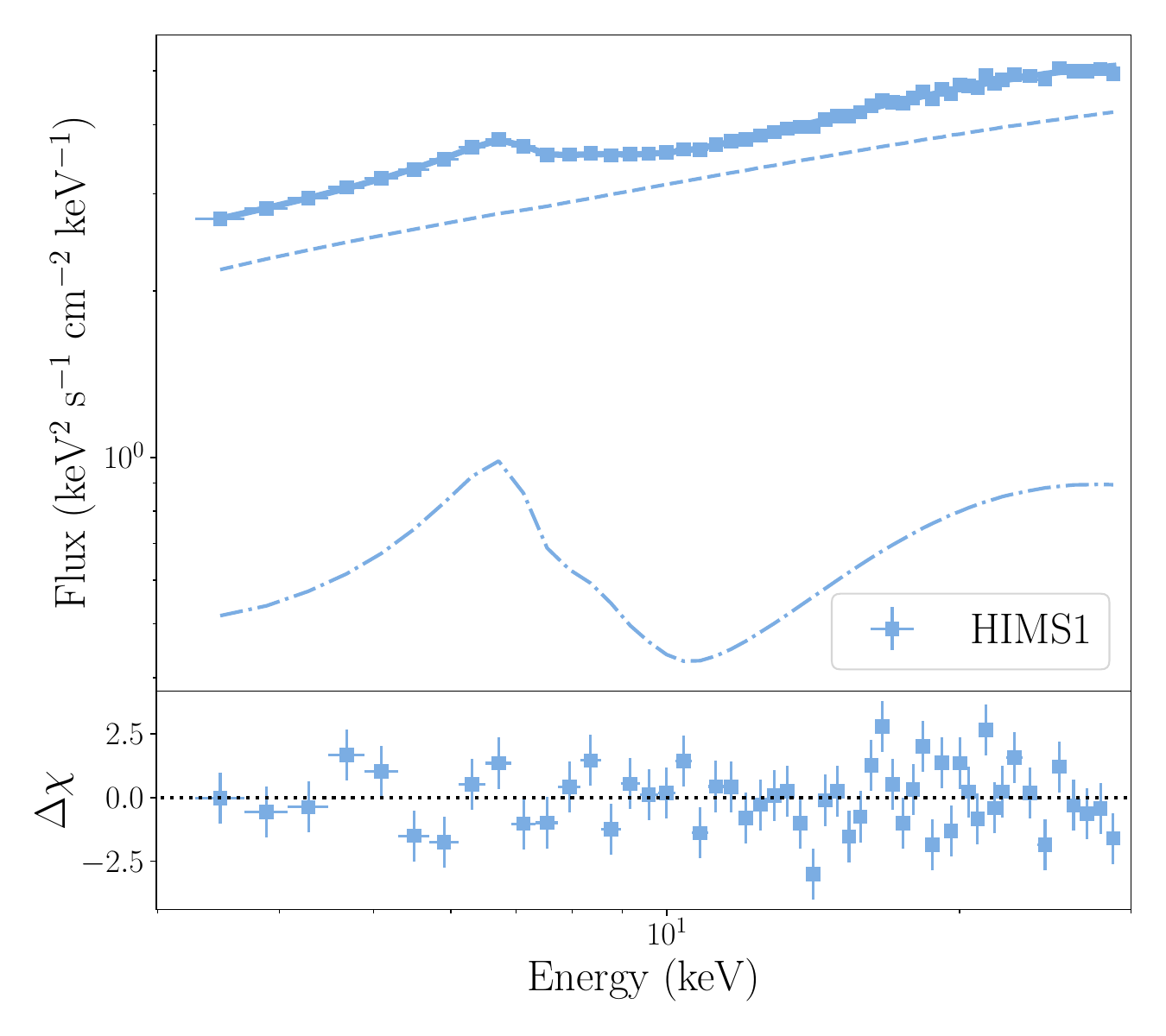}
    \includegraphics[width=0.49\columnwidth, trim={0.65cm 0.0cm 0.6cm 0.0cm},clip]{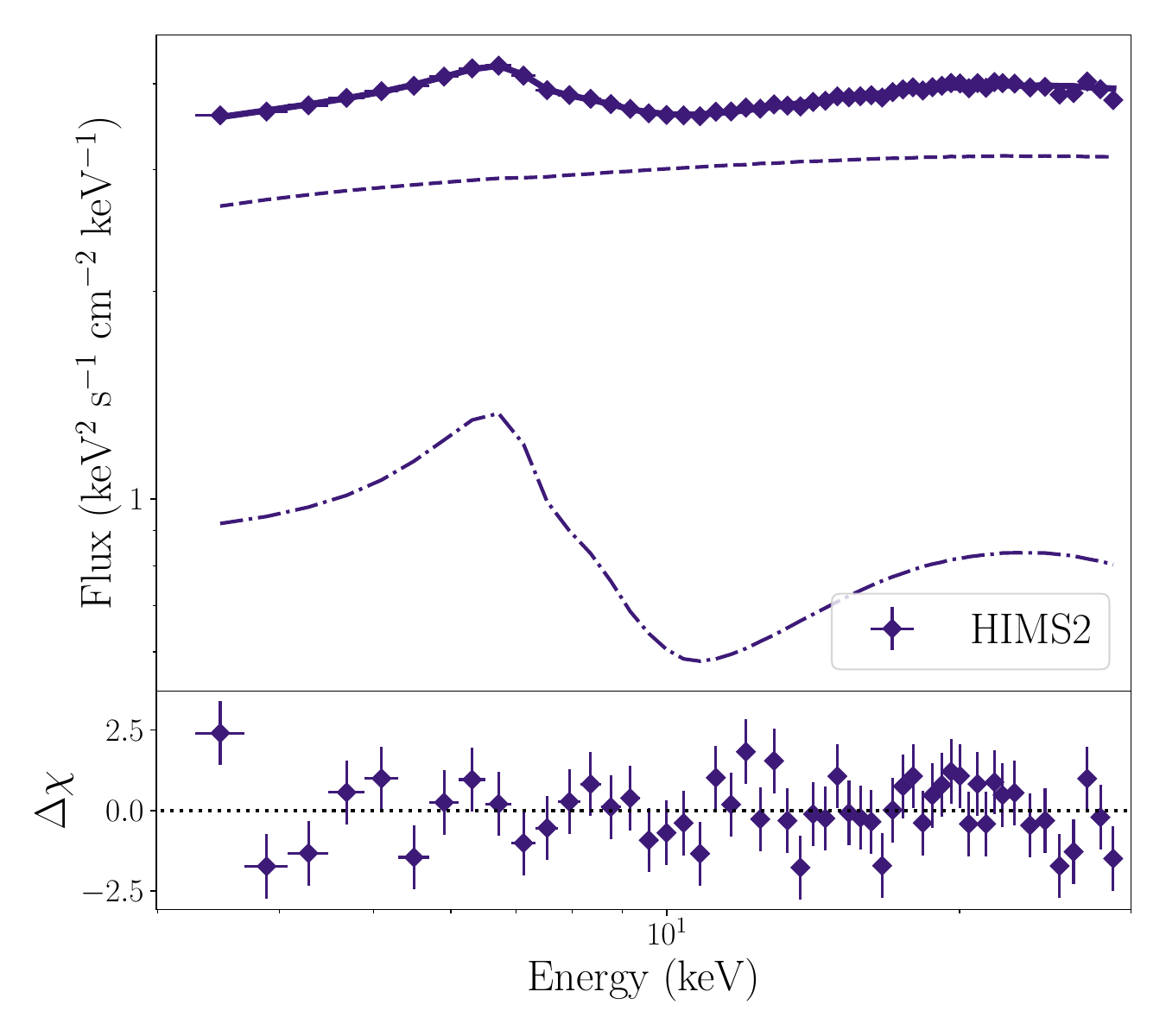}
    \caption{Time averaged energy spectra of the four epochs highlighted in Fig.\ref{fig:hardness_hue}. Dashed lines indicate the continuum, dash-dotted lines indicate the reflection, and continuous lines indicate the total model.}
    \label{fig:energyspectra}
\end{figure*}

\begin{figure*}[h]
    \centering
    \includegraphics[width=0.49\columnwidth]{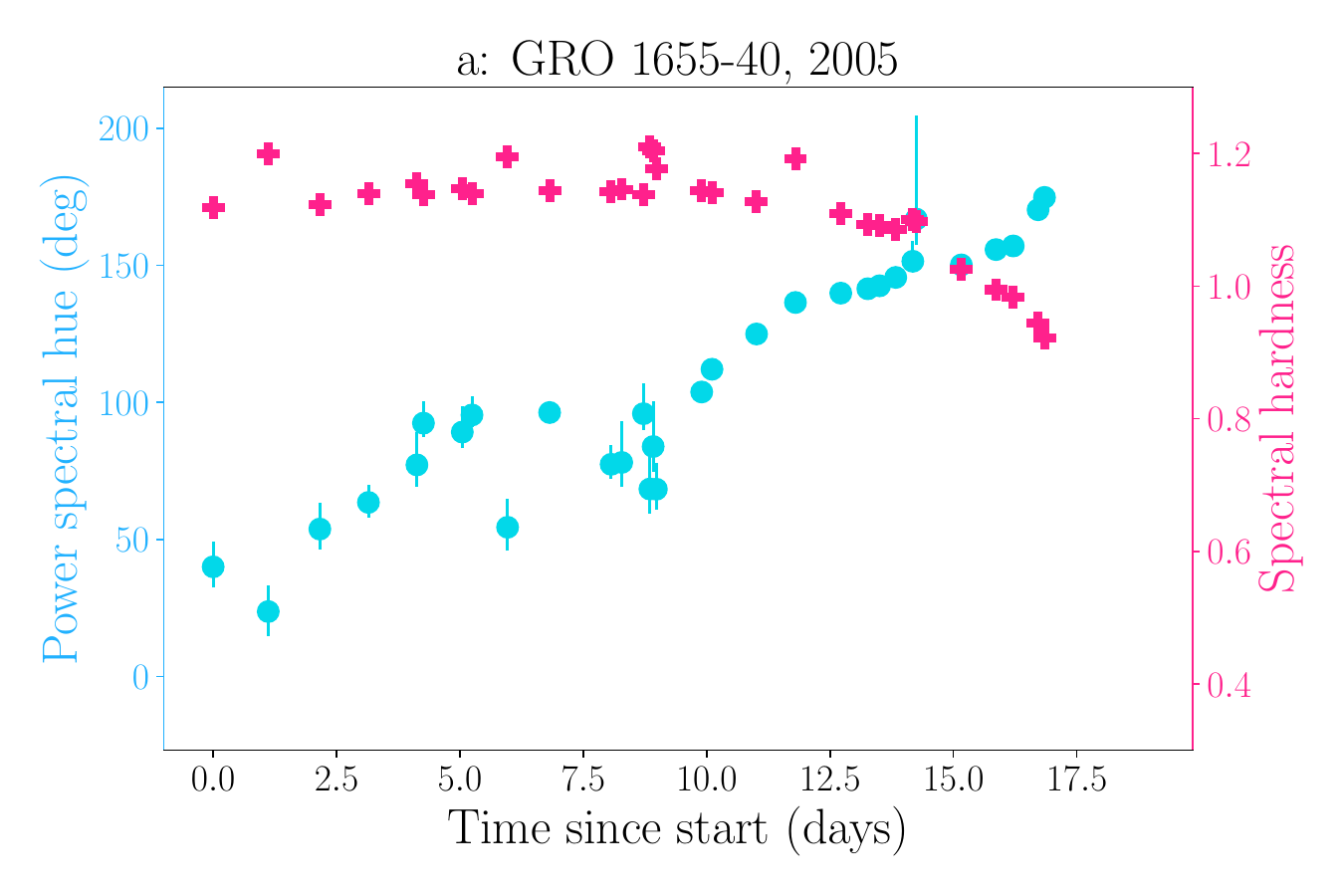}
    \includegraphics[width=0.49\columnwidth]{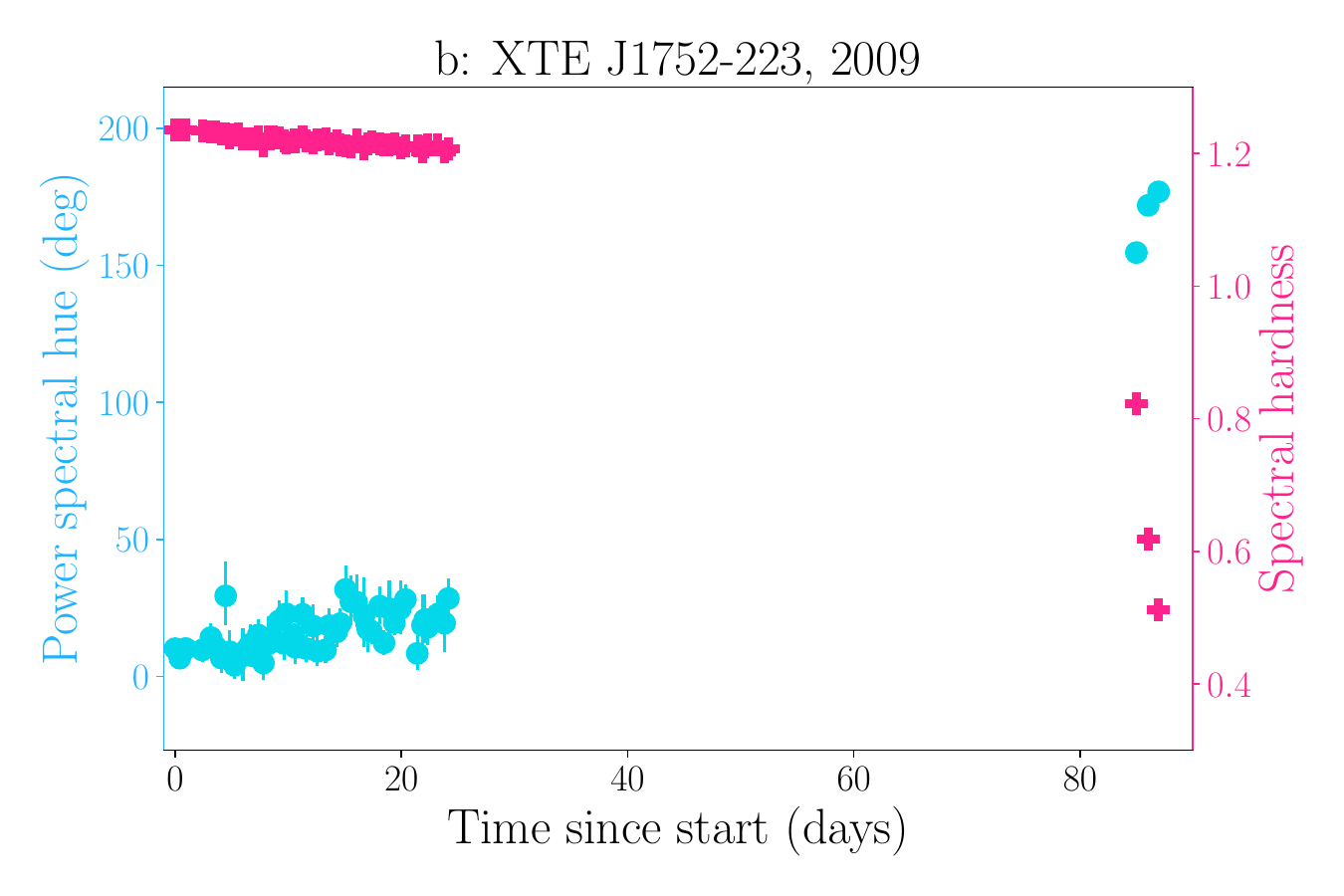}
    \includegraphics[width=0.49\columnwidth]{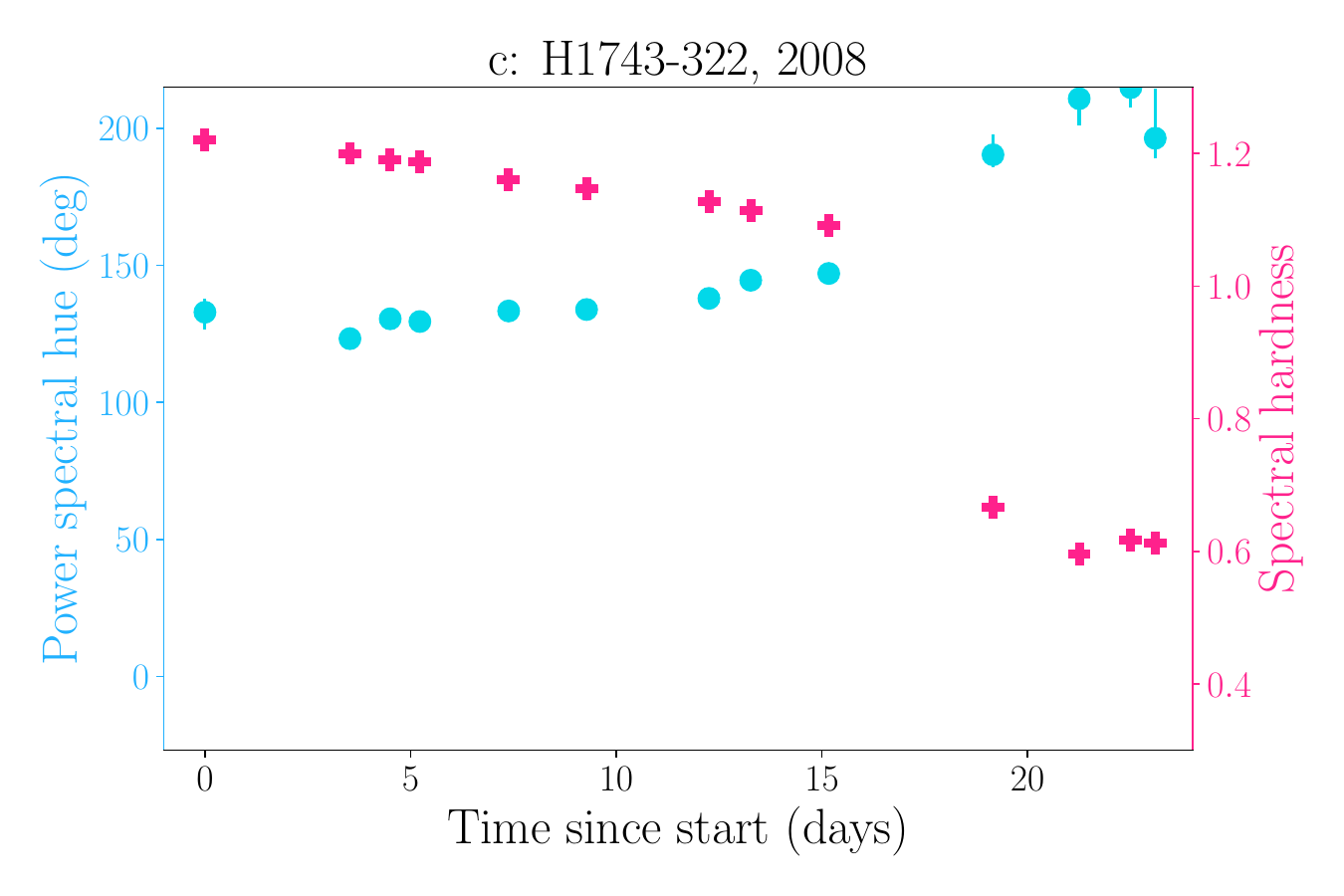}
    \includegraphics[width=0.49\columnwidth]{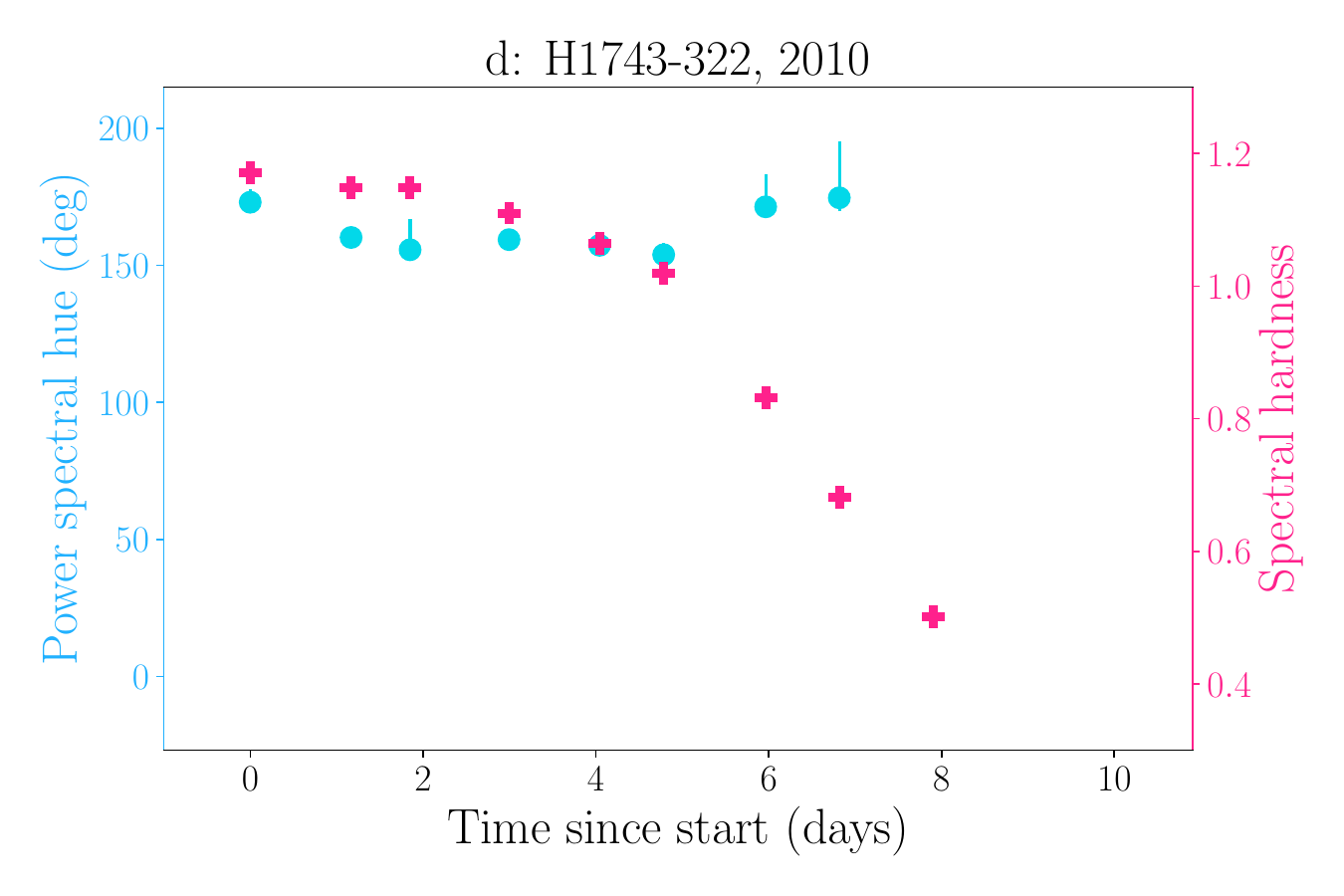}
    \includegraphics[width=0.49\columnwidth]{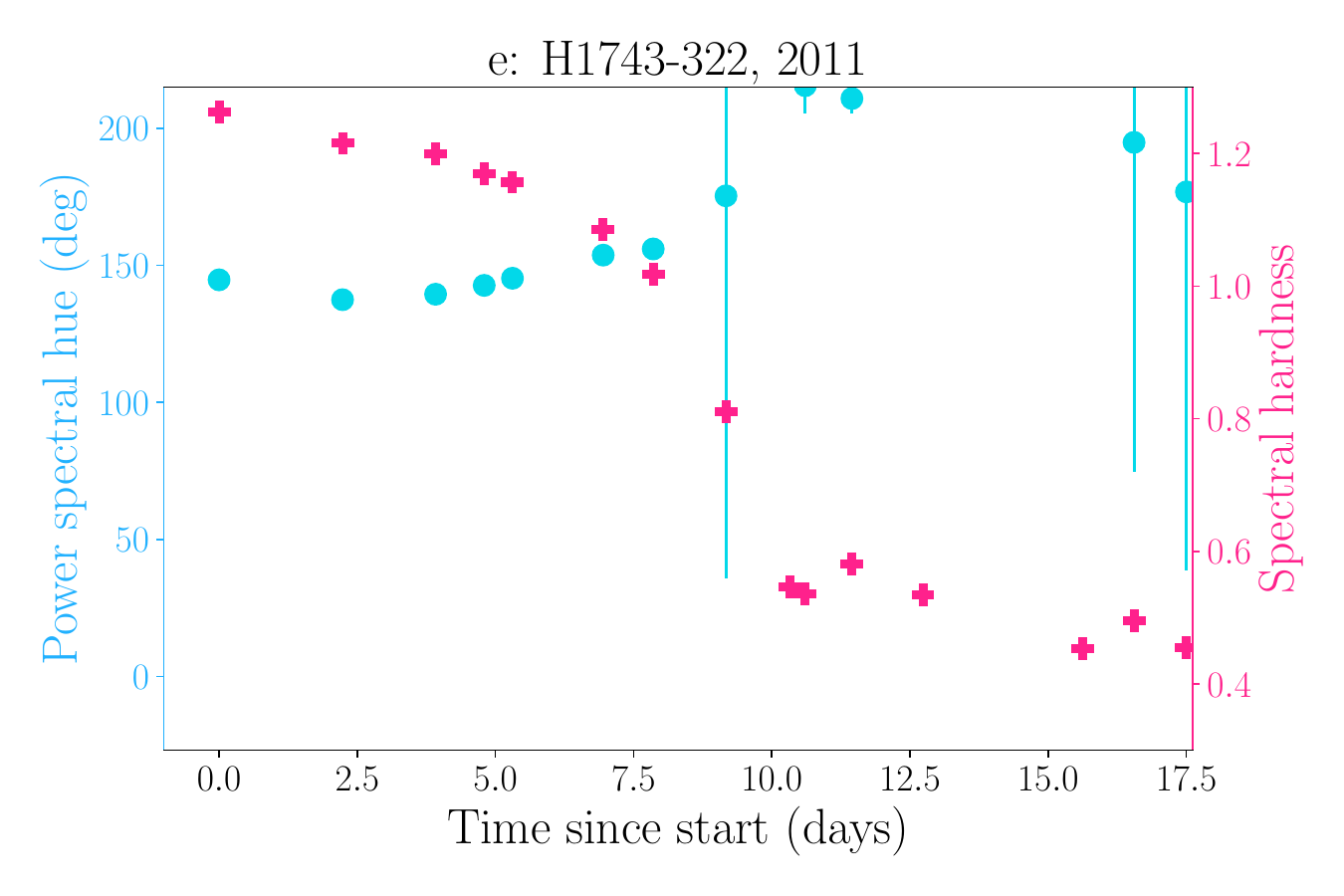}
    \caption{Evolution of the other outbursts analyzed in this work, which have more limited coverage than GX 339$-$4. Top left: the 2005 outburst of GRO 1655$-$40. Top right: the 2009 outburst of XTE J1752$-$223. Remaining panels: the 2008, 2010 and 2011 outbursts of H 1743-322. The color coding is identical to Fig.\ref{fig:hardness_hue}.}
    \label{fig:hardness_hue_all_1}
\end{figure*}

\begin{figure*}[h]
    \centering
    \includegraphics[width=0.49\columnwidth, trim={0.65cm 0.0cm 0.6cm 0.0cm},clip]{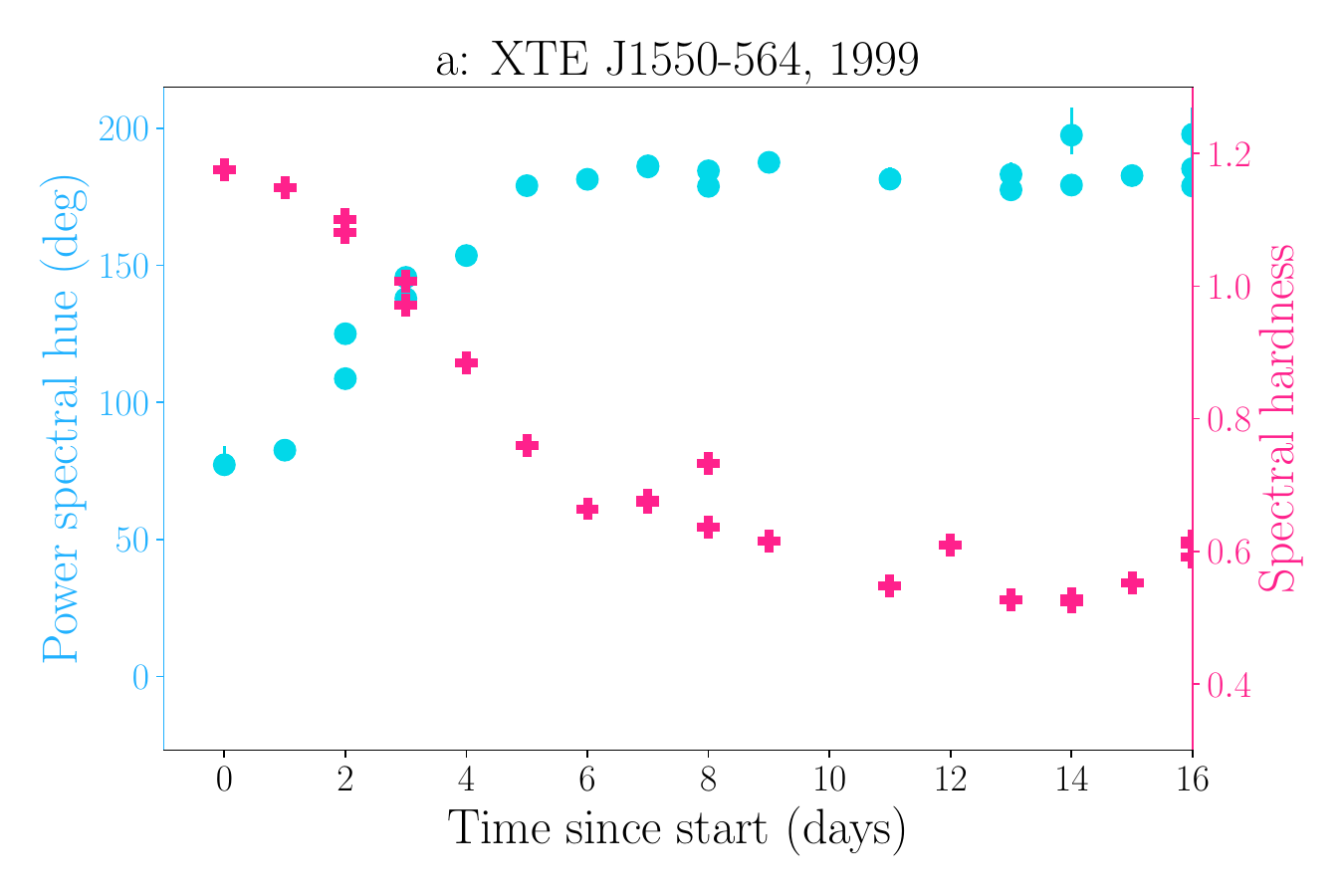}
    \includegraphics[width=0.49\columnwidth, trim={0.65cm 0.0cm 0.6cm 0.0cm},clip]{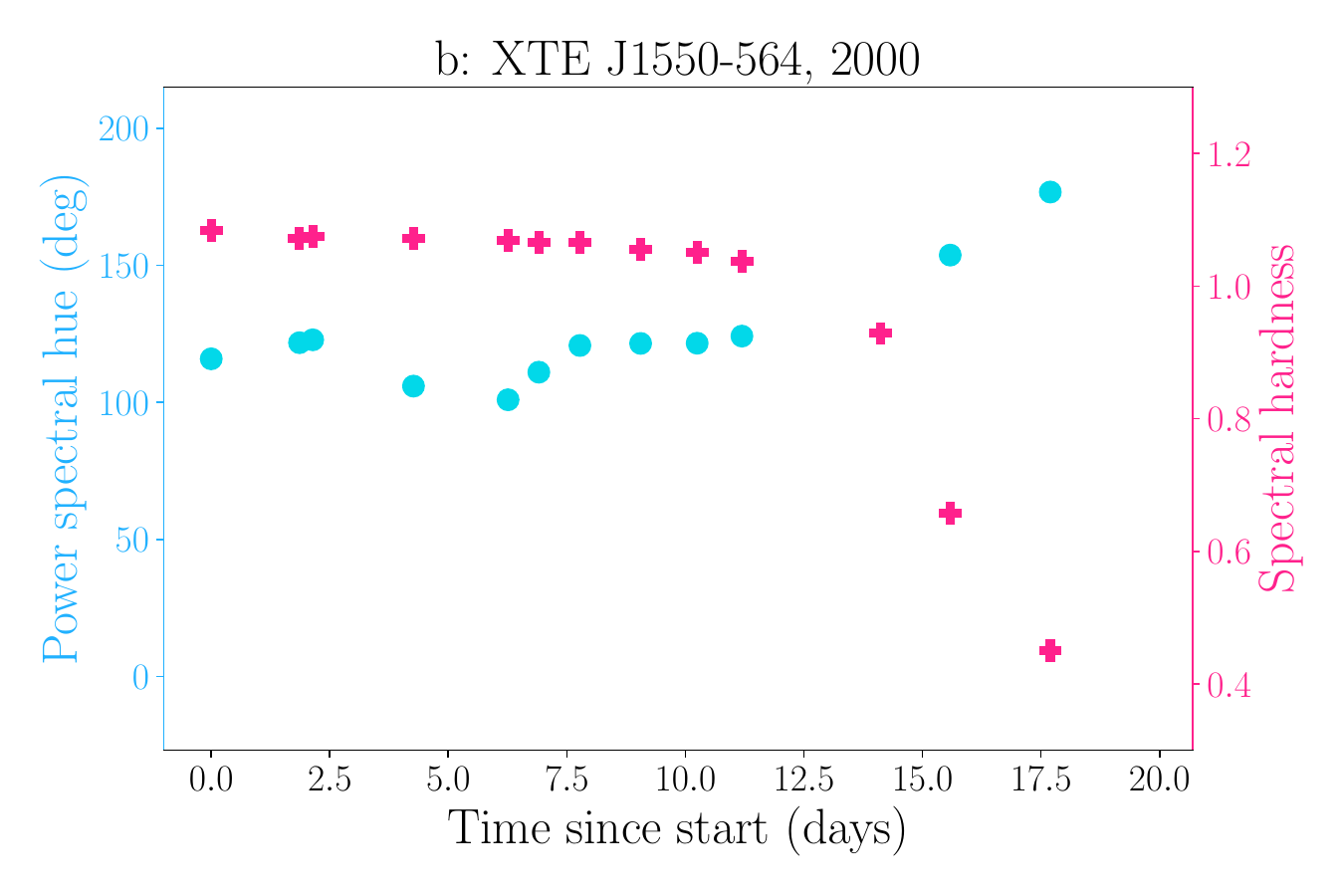}
    \includegraphics[width=0.49\columnwidth, trim={0.65cm 0.0cm 0.6cm 0.0cm},clip]{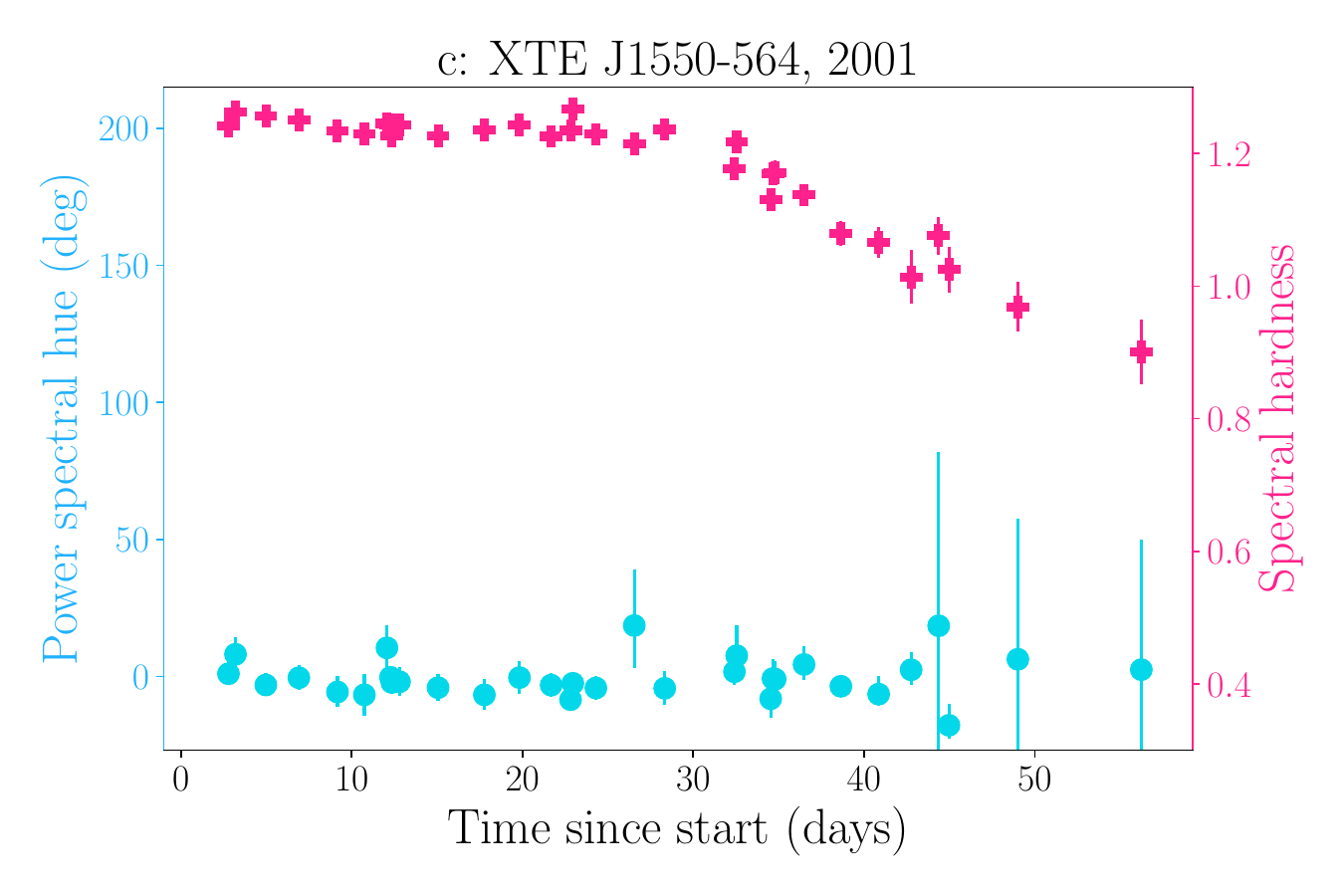}
    \includegraphics[width=0.49\columnwidth, trim={0.65cm 0.0cm 0.6cm 0.0cm},clip]{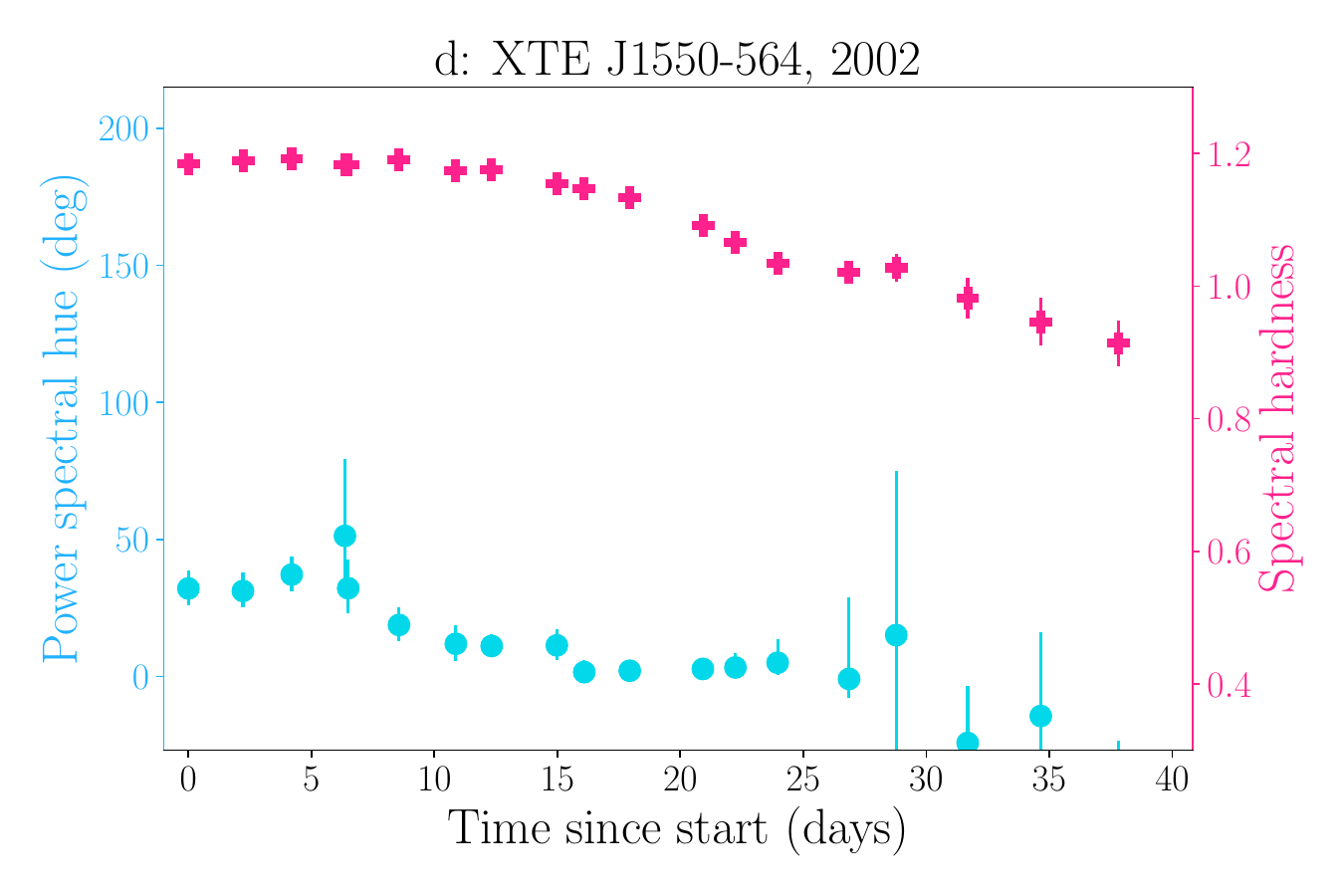}
    \includegraphics[width=0.49\columnwidth, trim={0.65cm 0.0cm 0.6cm 0.0cm},clip]{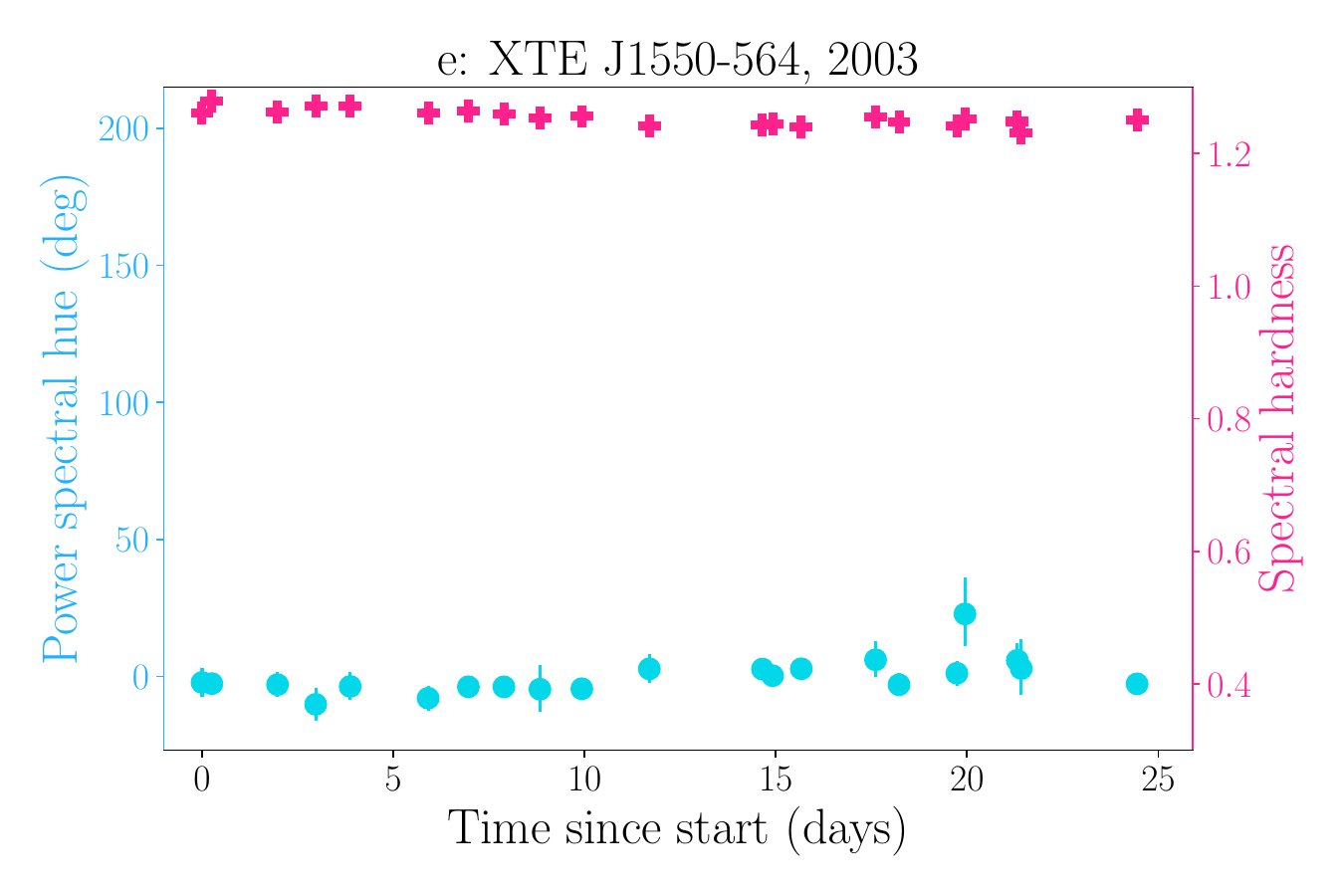}
    \caption{Evolution of all the outbursts XTE J1550$-$564; similarly to the previous panel, the coverage of the rise or state transition. The panels from left to right and top to bottom correspond to the 1999, 2000 full outbursts and 2001, 2002 and 2003 hard state outbursts. The color coding is identical to Fig.\ref{fig:hardness_hue}.}
    \label{fig:hardness_hue_all_2}
\end{figure*}

In this section we provide additional information about the data we analyzed. Tab.\ref{tab:outburst_date_1} and Tab.\ref{tab:outburst_date_2} summarize the properties of all the outbursts we analyzed. Tab.\ref{tab:PSD_fits} and \ref{tab:spec_fits} report the best-fitting values for the power spectra and time-averaged energy spectra, respectively. Fig.\ref{fig:energyspectra} shows the time-averaged spectra from the four observations discussed in the main text. Fig.\ref{fig:hardness_hue_1}, \ref{fig:hardness_hue_all_1} and \ref{fig:hardness_hue_all_2} show the evolution of the hue and hardness for all the outbursts analzyed in this work. We only modeled the evolution of the power spectral hue and spectral hardness in GX 339$-$4 due to the more regular behavior and better sampling of the rising hard state and state transition.

\begin{table}[h]
    \centering
    \begin{tabular}{|c|cccc|}
    \hline 
     Outburst year and type & Start (MJD) & End (MJD) & $t_{\rm trans, hue}$ (days) & $t_{\rm trans, hr}$ (days) \\ \hline
     GX 339$-$4 2002, Full & 52359 & 52784& $20^{+5}_{-5}$ & $40.08^{+0.26}_{-0.26}$ \\
     GX 339$-$4 2004, Full & 53044 & 53514 & $143^{+7}_{-7.3}$ & $180.0^{+0.25}_{-0.25}$ \\
     GX 339$-$4 2006, Hard State & 53769 & 53878 &  & \\
     GX 339$-$4 2007, Full & 54060 & 54385 & $64.3^{+0.7}_{-0.7}$ & $77.08^{+0.21}_{-0.21}$ \\
     GX 339$-$4 2008, Hard State & 54631 & 54758 &   & \\
     GX 339$-$4 2009, Hard State & 54884 & 55023 &   & \\
     GX 339$-$4 2010, Full & 55178 & 55632& $65.0^{+2.5}_{-2.5}$ & $87.11^{+0.17}_{-0.17}$ \\
     \hline
    \end{tabular}
    \caption{Properties of all the outbursts of GX 339-4 analyzed in this work. Start and end dates have been defined from \cite{Plant14} and \cite{Grebenev20}.}
    \label{tab:outburst_date_1}
\end{table}

\begin{table}[h]
    \centering
    \begin{tabular}{|c|ccc|}
    \hline 
     Source, outburst year and type & Start (MJD) & End (MJD) & Reference \\ \hline
     GRO 1655$-$40 2005, Full & 53420 & 53445 & \cite{Shaposhnikov07} \\
     XTE J1752$-$223 2009, Full & 55125 & 55350 & \cite{Shaposhnikov10}  \\
     H1743$-$322 2008, Full & 54732 & 54804 & \cite{Grebenev20} \\
     H1743$-$322 2010, Full & 55411 & 55471 & \cite{Grebenev20} \\
     H1743$-$322 2011, Full & 55656 & 55712 & \cite{Grebenev20} \\
     XTE J1550$-$564 1999, Full & 51063 & 51280 & \cite{Cui99}, \cite{Sobczak00} \\
     XTE J1550$-$564 2000, Full & 51635 & 51700 & \cite{Jain01} \\
     XTE J1550$-$564 2001, Hard State & 51900 & 52000 & \cite{Curran13} \\
     XTE J1550$-$564 2002, Hard State & 52250 & 52350 & \cite{Curran13} \\
     XTE J1550$-$564 2003, Hard State & 52700 & 52800 & \cite{Curran13} \\ 
     \hline
    \end{tabular}
    \caption{Properties of the additional outbursts analyzed in this work. }
    \label{tab:outburst_date_2}
\end{table}

\begin{table}[h]
    \centering
    \begin{tabular}{|c|cccc|} 
    \hline
     & 92052-07-03-01 & 92052-07-06-01 & 92035-01-02-03 & 92035-01-03-00\\
    \hline
    Lorentz 1 $\sigma$ (Hz) & $0.13^{+0.07}_{-0.05}$ & $0.21^{+0.03}_{-0.03}$ & $0.75^{+0.05}_{-0.05}$ & $1.27^{+0.14}_{-0.16}$ \\
    Lorentz 1 $K$ ($\rm{rms}\,Hz^{-1}$) & $5.5^{+1.0}_{-1.2}\times 10^{-2}$ & $4.45^{+0.42}_{-0.42}\times 10^{-2}$ & $2.82^{+0.15}_{-0.15}\times 10^{-2}$ & $6.18^{+0.86}_{-0.39}\times 10^{-3}$ \\
    Lorentz 2 $f_c$ (Hz) & $<2.4$ & $0.3^{+0.3}_{-0.3}$ & $2.24^{+0.28}_{-0.30}$ &  $7.76^{+0.09}_{-0.17}\times10^{-3}$ \\
    Lorentz 2 $\sigma$ (Hz) & $2.34^{+0.31}_{-0.25}$ & $4.6^{+0.3}_{-0.3}$ & $7.22^{+0.26}_{-0.22}$ &  $9.8^{+1.5}_{-1.5}\times10^{-2}$ \\
    Lorentz 2 $K$ ($\rm{rms}\,Hz^{-1}$) & $8.4^{+0.6}_{-0.6}\times 10^{-2}$ & $3.6^{+0.7}_{-0.5}\times 10^{-2}$ & $1.38^{+0.07}_{-0.07}\times 10^{-2}$ &  $1.28^{+0.04}_{-0.09}\times10^{-2}$  \\
    Lorentz 3 $f_c$ (Hz) & // & // & $0.36^{+0.06}_{-0.06}$ & $1.127^{+0.011}_{-0.019}$ \\
    Lorentz 3 $\sigma$ (Hz) & // & // & $5.4^{+1.1}_{-1.3}\times 10^{-2}$ &  $9.8^{+1.5}_{-1.5}\times10^{-2}$  \\
    Lorentz 3 $K$ ($\rm{rms}\,Hz^{-1}$) & // & // & $7.5^{+1.2}_{-1.2}\times 10^{-2}$ &  $1.24^{+0.02}_{-0.2}\times10^{-3}$  \\
    Lorentz 4 $f_c$ (Hz) & // & // & // &  $2.281^{+0.007}_{-0.012}$  \\
    Lorentz 4 $\sigma$ (Hz) & // & // & // &  $1.67^{+0.01}_{-0.75}\times10^{-1}$  \\
    Lorentz 4 $K$ ($\rm{rms}\,Hz^{-1}$) & // & // & // & $3.82^{+0.28}_{-0.34}\times10^{-3}$ \\
    \hline 
    $\Delta \chi^{2}/\rm{d.o.f.}$ & $53.52/72$ & $82.04/72$ & $115.32/69$ & $78.30/66$ \\
    \hline
    \end{tabular}
    \\
    \caption{Best fit of the power spectra for each of the four epochs highlighted in Fig.\ref{fig:hardness_hue}. The first Lorentzian was fixed to be zero-centered fits. Limits shown at 90\% confidence.}
    \label{tab:PSD_fits}
\end{table}

\begin{table}[h]
    \centering
    \begin{tabular}{|c|cccc|} 
    \hline
     & 92052-07-03-01 & 92052-07-06-01 & 92035-01-02-03 & 92035-01-03-00\\
    \hline
    $nH$ (cm$^{-2}$) & \multicolumn{4}{c|}{$0.7^{*}$} \\ 
    $i$ (deg) & \multicolumn{4}{c|}{$32.5^{+1.9}_{-2.5}$} \\ 
    $q$  & $3^{*}$ & $3.0^{+1.1}_{-0.6}$ & $2.9^{+0.4}_{-0.5}$ & $2.32^{+0.21}_{-0.14}$ \\
    $R_{\rm in}$ (R$_g$) & $6^{+8}_{-4}$ & $4.5^{+2.8}_{-2.4}$ & $4.9^{+1.6}_{-3.0}$ & $2.3^{+2.0}_{-1.0}$ \\
    $\log(\xi)$ & $1.9^{+0.6}_{-0.4}$ & $3.17^{+0.05}_{-0.05}$ & $3.49^{+0.09}_{-0.13}$ &  $3.76^{+0.11}_{-0.09}$ \\
    $A_{\rm Fe}$  & \multicolumn{4}{c|}{$4.3^{+0.6}_{-0.5}$}\\
    $\Gamma$ & $1.59^{+0.03}_{-0.03}$ & $1.653^{+0.012}_{-0.012}$ & $1.732^{+0.009}_{-0.007}$ &  $1.923^{+0.009}_{-0.007}$  \\
    $kT_{\rm e}$ (keV) & $300^{*}$ & $300^{*}$ & $26{+5}_{-4}$ & $38^{+8}_{-4}$ \\
    $\rm{Norm_{relxill}}$ &  $0.9^{+0.2}_{-0.4}\times 10^{-3}$ & $8.0^{+1.0}_{-1.0}\times 10^{-3}$ & $8.5^{+1.0}_{-1.0}\times 10^{-3}$ &  $4.1^{+0.4}_{-0.4}$ \\
    $\rm{Norm_{nthcomp}}$ & $9.2^{+0.5}_{-0.7}\times 10^{-2}$ & $0.74^{+0.02}_{-0.02}$ & $1.62^{+0.08}_{-0.07}$ & $2.52^{+0.13}_{-0.18}$ \\
    \hline 
    $\Delta \chi^{2}/\rm{\# bins}$ & $33.60/53$ & $53.44/54$ & $78.11/54$ & $50.82/54$ \\
    \hline
    \end{tabular}
    \\
    $^*$: parameter frozen
    \caption{Best fit of the  time-averaged energy spectra for each of the four epochs highlighted in Fig.\ref{fig:hardness_hue}. Limits shown at 90\% confidence. The total fit statistic is $\Delta \chi^{2}/\rm{d.o.f.}=215.98/188$.}
    \label{tab:spec_fits}
\end{table}

\bibliography{references}{}
\bibliographystyle{aasjournal}



\end{document}